\documentstyle[12pt]{article}
\advance \topmargin by -\headheight
\advance \topmargin by -\headsep

\evensidemargin \oddsidemargin
\begin{document}

\topmargin -.6in

\def\rh{{\hat \rho}}
\def\alie{{\hat{\cal G}}}
\newcommand{\sect}[1]{\setcounter{equation}{0}\section{#1}}
\renewcommand{\theequation}{\thesection.\arabic{equation}}

\def\rf#1{(\ref{eq:#1})}
\def\lab#1{\label{eq:#1}}
\def\nonu{\nonumber}
\def\br{\begin{eqnarray}}
\def\er{\end{eqnarray}}
\def\be{\begin{equation}}
\def\ee{\end{equation}}
\def\eq{\!\!\!\! &=& \!\!\!\! }
\def\foot#1{\footnotemark\footnotetext{#1}}
\def\lb{\lbrack}
\def\rb{\rbrack}
\def\llangle{\left\langle}
\def\rrangle{\right\rangle}
\def\blangle{\Bigl\langle}
\def\brangle{\Bigr\rangle}
\def\llbrack{\left\lbrack}
\def\rrbrack{\right\rbrack}
\def\lcurl{\left\{}
\def\rcurl{\right\}}
\def\({\left(}
\def\){\right)}
\newcommand{\nit}{\noindent}
\newcommand{\ct}[1]{\cite{#1}}
\newcommand{\bi}[1]{\bibitem{#1}}
\def\lskip{\vskip\baselineskip\vskip-\parskip\noindent}
\relax

\def\tr{\mathop{\rm tr}}
\def\Tr{\mathop{\rm Tr}}
\def\v{\vert}
\def\bv{\bigm\vert}
\def\Bgv{\;\Bigg\vert}
\def\bgv{\bigg\vert}
\newcommand\partder[2]{{{\partial {#1}}\over{\partial {#2}}}}
\newcommand\funcder[2]{{{\delta {#1}}\over{\delta {#2}}}}
\newcommand\Bil[2]{\Bigl\langle {#1} \Bigg\vert {#2} \Bigr\rangle}  
\newcommand\bil[2]{\left\langle {#1} \bigg\vert {#2} \right\rangle} 
\newcommand\me[2]{\left\langle {#1}\bv {#2} \right\rangle} 
\newcommand\sbr[2]{\left\lbrack\,{#1}\, ,\,{#2}\,\right\rbrack}
\newcommand\pbr[2]{\{\,{#1}\, ,\,{#2}\,\}}
\newcommand\pbbr[2]{\lcurl\,{#1}\, ,\,{#2}\,\rcurl}
%
\def\a{\alpha}
\def\at{{\tilde A}^R}
\def\atc{{\tilde {\cal A}}^R}
\def\atcm#1{{\tilde {\cal A}}^{(R,#1)}}
\def\b{\beta}
\def\dc{{\cal D}}
\def\d{\delta}
\def\D{\Delta}
\def\eps{\epsilon}
\def\vareps{\varepsilon}
\def\g{\gamma}
\def\G{\Gamma}
\def\grad{\nabla}
\def\h{{1\over 2}}
\def\l{\lambda}
\def\L{\Lambda}
\def\m{\mu}
\def\n{\nu}
\def\o{\over}
\def\om{\omega}
\def\O{\Omega}
\def\p{\phi}
\def\P{\Phi}
\def\pa{\partial}
\def\pr{\prime}
\def\pt{{\tilde \Phi}}
\def\qs{Q_{\bf s}}
\def\ra{\rightarrow}
\def\s{\sigma}
\def\S{\Sigma}
\def\t{\tau}
\def\th{\theta}
\def\Th{\Theta}
\def\tpp{\Theta_{+}}
\def\tmm{\Theta_{-}}
\def\tpg{\Theta_{+}^{>}}
\def\tms{\Theta_{-}^{<}}
\def\tp0{\Theta_{+}^{(0)}}
\def\tm0{\Theta_{-}^{(0)}}
\def\ti{\tilde}
\def\wti{\widetilde}
\def\jc{J^C}
\def\bj{{\bar J}}
\def\sj{{\jmath}}
\def\bsj{{\bar \jmath}}
\def\bp{{\bar \p}}
\def\vp{\varphi}
\def\vt{{\tilde \varphi}}
\def\faa{Fa\'a di Bruno~}
\def\ca{{\cal A}}
\def\cb{{\cal B}}
\def\ce{{\cal E}}
\def\cg{{\cal G}}
\def\cgh{{\hat {\cal G}}}
\def\ch{{\cal H}}
\def\chh{{\hat {\cal H}}}
\def\cl{{\cal L}}
\def\cm{{\cal M}}
\def\cn{{\cal N}}
\newcommand\sumi[1]{\sum_{#1}^{\infty}}   
\newcommand\fourmat[4]{\left(\begin{array}{cc}  
{#1} & {#2} \\ {#3} & {#4} \end{array} \right)}

%
\def\lie{{\cal G}}
\def\kmlie{{\hat{\cal G}}}
\def\dlie{{\cal G}^{\ast}}
\def\elie{{\widetilde \lie}}
\def\edlie{{\elie}^{\ast}}
\def\hlie{{\cal H}}
\def\flie{{\cal F}}
\def\wlie{{\widetilde \lie}}
\def\f#1#2#3 {f^{#1#2}_{#3}}
\def\winf{{\sf w_\infty}}
\def\win1{{\sf w_{1+\infty}}}
\def\hwinf{{\sf {\hat w}_{\infty}}}
\def\Winf{{\sf W_\infty}}
\def\Win1{{\sf W_{1+\infty}}}
\def\hWinf{{\sf {\hat W}_{\infty}}}
\def\Rm#1#2{r(\vec{#1},\vec{#2})}          
\def\OR#1{{\cal O}(R_{#1})}           
\def\ORti{{\cal O}({\widetilde R})}           
\def\AdR#1{Ad_{R_{#1}}}              
\def\dAdR#1{Ad_{R_{#1}^{\ast}}}      
\def\adR#1{ad_{R_{#1}^{\ast}}}       
\def\KP{${\rm \, KP\,}$}                 
\def\KPl{${\rm \,KP}_{\ell}\,$}         
\def\KPo{${\rm \,KP}_{\ell = 0}\,$}         
\def\mKPa{${\rm \,KP}_{\ell = 1}\,$}    
\def\mKPb{${\rm \,KP}_{\ell = 2}\,$}    
%
\def\rlx{\relax\leavevmode}
\def\inbar{\vrule height1.5ex width.4pt depth0pt}
\def\IZ{\rlx\hbox{\sf Z\kern-.4em Z}}
\def\IR{\rlx\hbox{\rm I\kern-.18em R}}
\def\IC{\rlx\hbox{\,$\inbar\kern-.3em{\rm C}$}}
\def\IN{\rlx\hbox{\rm I\kern-.18em N}}
\def\IO{\rlx\hbox{\,$\inbar\kern-.3em{\rm O}$}}
\def\IP{\rlx\hbox{\rm I\kern-.18em P}}
\def\IQ{\rlx\hbox{\,$\inbar\kern-.3em{\rm Q}$}}
\def\IF{\rlx\hbox{\rm I\kern-.18em F}}
\def\IG{\rlx\hbox{\,$\inbar\kern-.3em{\rm G}$}}
\def\IH{\rlx\hbox{\rm I\kern-.18em H}}
\def\II{\rlx\hbox{\rm I\kern-.18em I}}
\def\IK{\rlx\hbox{\rm I\kern-.18em K}}
\def\IL{\rlx\hbox{\rm I\kern-.18em L}}
\def\one{\hbox{{1}\kern-.25em\hbox{l}}}
\def\0#1{\relax\ifmmode\mathaccent"7017{#1}%
B        \else\accent23#1\relax\fi}
\def\omz{\0 \omega}
%
\def\ltimes{\mathrel{\vrule height1ex}\joinrel\mathrel\times}
\def\rtimes{\mathrel\times\joinrel\mathrel{\vrule height1ex}}
%
\def\mark{\noindent{\bf Remark.}\quad}
\def\prop{\noindent{\bf Proposition.}\quad}
\def\theor{\noindent{\bf Theorem.}\quad}
\def\name{\noindent{\bf Definition.}\quad}
\def\exam{\noindent{\bf Example.}\quad}
\def\proof{\noindent{\bf Proof.}\quad}
%
%
\def\PRL#1#2#3{{\sl Phys. Rev. Lett.} {\bf#1} (#2) #3}
\def\NPB#1#2#3{{\sl Nucl. Phys.} {\bf B#1} (#2) #3}
\def\NPBFS#1#2#3#4{{\sl Nucl. Phys.} {\bf B#2} [FS#1] (#3) #4}
\def\CMP#1#2#3{{\sl Commun. Math. Phys.} {\bf #1} (#2) #3}
\def\PRD#1#2#3{{\sl Phys. Rev.} {\bf D#1} (#2) #3}
\def\PLA#1#2#3{{\sl Phys. Lett.} {\bf #1A} (#2) #3}
\def\PLB#1#2#3{{\sl Phys. Lett.} {\bf #1B} (#2) #3}
\def\JMP#1#2#3{{\sl J. Math. Phys.} {\bf #1} (#2) #3}
\def\PTP#1#2#3{{\sl Prog. Theor. Phys.} {\bf #1} (#2) #3}
\def\SPTP#1#2#3{{\sl Suppl. Prog. Theor. Phys.} {\bf #1} (#2) #3}
\def\AoP#1#2#3{{\sl Ann. of Phys.} {\bf #1} (#2) #3}
\def\PNAS#1#2#3{{\sl Proc. Natl. Acad. Sci. USA} {\bf #1} (#2) #3}
\def\RMP#1#2#3{{\sl Rev. Mod. Phys.} {\bf #1} (#2) #3}
\def\PR#1#2#3{{\sl Phys. Reports} {\bf #1} (#2) #3}
\def\AoM#1#2#3{{\sl Ann. of Math.} {\bf #1} (#2) #3}
\def\UMN#1#2#3{{\sl Usp. Mat. Nauk} {\bf #1} (#2) #3}
\def\FAP#1#2#3{{\sl Funkt. Anal. Prilozheniya} {\bf #1} (#2) #3}
\def\FAaIA#1#2#3{{\sl Functional Analysis and Its Application} {\bf #1} (#2)
#3}
\def\BAMS#1#2#3{{\sl Bull. Am. Math. Soc.} {\bf #1} (#2) #3}
\def\TAMS#1#2#3{{\sl Trans. Am. Math. Soc.} {\bf #1} (#2) #3}
\def\InvM#1#2#3{{\sl Invent. Math.} {\bf #1} (#2) #3}
\def\LMP#1#2#3{{\sl Letters in Math. Phys.} {\bf #1} (#2) #3}
\def\IJMPA#1#2#3{{\sl Int. J. Mod. Phys.} {\bf A#1} (#2) #3}
\def\AdM#1#2#3{{\sl Advances in Math.} {\bf #1} (#2) #3}
\def\RMaP#1#2#3{{\sl Reports on Math. Phys.} {\bf #1} (#2) #3}
\def\IJM#1#2#3{{\sl Ill. J. Math.} {\bf #1} (#2) #3}
\def\APP#1#2#3{{\sl Acta Phys. Polon.} {\bf #1} (#2) #3}
\def\TMP#1#2#3{{\sl Theor. Mat. Phys.} {\bf #1} (#2) #3}
\def\JPA#1#2#3{{\sl J. Physics} {\bf A#1} (#2) #3}
\def\JSM#1#2#3{{\sl J. Soviet Math.} {\bf #1} (#2) #3}
\def\MPLA#1#2#3{{\sl Mod. Phys. Lett.} {\bf A#1} (#2) #3}
\def\JETP#1#2#3{{\sl Sov. Phys. JETP} {\bf #1} (#2) #3}
\def\JETPL#1#2#3{{\sl  Sov. Phys. JETP Lett.} {\bf #1} (#2) #3}
\def\PHSA#1#2#3{{\sl Physica} {\bf A#1} (#2) #3}
\def\PHSD#1#2#3{{\sl Physica} {\bf D#1} (#2) #3}
\def\PJA#1#2#3{{\sl Proc. Japan. Acad} {\bf #1A} (#2) #3}
\def\JPSJ#1#2#3{{\sl J. Phys. Soc. Japan} {\bf #1} (#2) #3}

\begin{titlepage}
\vspace*{-1cm}
\noindent
November, 1994 \hfill{US-FT/21-94}\\
\phantom{bla}
\hfill{IFT-P/002/95} \\
\phantom{bla}
\hfill{hep-th/9412127}
\\
\vskip 2cm
\begin{center}
{\large\bf Solitons, Tau-functions and Hamiltonian Reduction for}
\end{center}
\begin{center}
{\large\bf Non-Abelian Conformal Affine Toda Theories}
\end{center}

\normalsize
\vskip 1.5cm

\begin{center}
L.A. Ferreira\footnotemark
{\footnotetext{ferreira@gaes.usc.es \newline
\indent $\;\;\;$On leave from Instituto
de F\'\i sica Te\'orica, IFT-UNESP - S\~ao Paulo-SP - Brazil.}},
J. Luis Miramontes\footnotemark
{\footnotetext{miramontes@gaes.usc.es}} and
Joaqu\'\i n S\'anchez Guill\'en\footnotemark
{\footnotetext{joaquin@gaes.usc.es}}

\par \vskip .1in \noindent
Departamento de F\'\i sica de Part\'\i culas,\\
Facultad de F\'\i sica\\
Universidad de Santiago\\
E-15706 Santiago de Compostela, Spain

\par \vskip .3in

\end{center}
\vfill
\begin{center}
{\large {\bf ABSTRACT}}\\
\end{center}
\par \vskip .3in

\noindent
We consider the Hamiltonian reduction of the two-loop
Wess-Zumino-Novikov-Witten model (WZNW) based on an untwisted affine Kac-Moody
algebra $\cgh$. The resulting reduced models, called
{\em Generalized Non-Abelian Conformal Affine Toda (G-CAT)}, are conformally
invariant and a wide class of them possesses soliton solutions; these models
constitute non-abelian generalizations of the Conformal Affine Toda models.
Their general solution is constructed by the Leznov-Saveliev method. Moreover,
the dressing transformations leading to the solutions in the orbit of the
vacuum are considered in detail, as well as the $\tau$-functions, which are
defined for any integrable highest weight representation of $\cgh$,
irrespectively of its particular realization. When
the conformal symmetry is spontaneously broken, the G-CAT model becomes a
generalized Affine Toda model, whose soliton solutions are
constructed. Their masses are obtained exploring the spontaneous breakdown
of the conformal symmetry, and their relation to the fundamental particle
masses is discussed.

\vskip 1.5cm

\end{titlepage}

\sect{Introduction}
\label{sec:intro}

\ \indent
The study of classical and quantum non-linear integrable models in
1+1~dimensions is of great interest in High Energy Physics, where such
models have been used as laboratories to develop methods to explore
the non-linear perturbative aspects of gauge theories, gravity and
string theory. In particular, they could help in understanding some
stable classical solutions, like monopoles, which must have an
important role in the quantum theory, and which cannot be understood by
the existing methods.

In this paper, we construct the Generalized Non-Abelian Conformal
Affine Toda models, the G-CAT models, which are a class of generalizations of
the well known non-abelian Affine Toda Field theories that, being
integrable and conformally invariant, provide one of the simplest example of
mass generation through spontaneous symmetry breaking. To be precise, within
these models, the mass of the solitons of the Affine Toda Theories can be
understood in terms of a Higgs-like mechanism associated to the spontaneous
breakdown of the conformal symmetry. Even more, within this approach, it is
possible to put in one-to-one correspondence the solitons of the
non-abelian Affine Toda models with its {\it massive} fundamental particles;
in particular, their masses are proportional.

Within the integrable models in 1+1~dimensions, the investigation of
the different Toda Field Theories has recently received a lot of
attention. According to their underlying algebraic structure, they can
be divided into three categories; each one exhibiting nice
characteristic properties. First, associated to the finite simple Lie
algebras, there are the Conformal Toda models, which are conformally
invariant 1+1 field theories. Even more, they permit the construction
of extensions of the Virasoro algebra including higher spin
generators, namely W-algebras. The second class of theories are the
Affine Toda models, based on loop algebras, which can be regarded as a
perturbed Conformal Toda model where the conformal symmetry is broken
by the perturbation while the integrability is preserved~\ct{perturb}. One of
their main properties is that they possess soliton solutions. These
two classes of models are called abelian or non-abelian
referring to whether their fields live on an abelian or
non-abelian group~\ct{misha,misha2,gervais,und}. Finally, the
conformal symmetry can be restored in the abelian Affine Toda models
just by adding two extra fields which do not modify the dynamics of the
original model; one of these fields is a connection whose only role is
to implement the conformal invariance. These are the so called
Conformal Affine Toda models~\ct{AFGZ,bb}, and they are based on a full
Kac-Moody algebra; moreover, they are integrable~\ct{charges}, and
have soliton solutions~\ct{acfgz}. In fact, many properties of the
Affine Toda models can be more easily understood by considering them as
the Conformal Affine Toda models with the conformal symmetry
spontaneously broken. All these Toda models can be obtained via
Hamiltonian reductions of WZNW models~\ct{dublin,AFGZ},whilst their
one dimensional versions can be obtained from free motion on symmetric
spaces~\ct{perel,fo}.

The G-CAT models consist on a WZNW model where the field takes values in a
finite ---not necessarily semisimple--- Lie group $G_0$, and where the
symmetries of left and right translations by elements of $G_0$ are broken by a
term of the form
\be
\int \, d^2 x \Tr \( \Lambda_{-l} B \Lambda_{l} B^{-1} \),\qquad
B\in G_0
\lab{breakterm}
\ee
where $\Lambda_{\pm l}$ are constant elements of subspaces of grade $\pm l$
of a Kac-Moody algebra $\cgh$; the grades corresponding to a gradation of
$\cgh$ where the subspace of zero grade is the Lie algebra of $G_0$. Notice
that we allow the central term of $\cgh$ and the grading operator to be
generators of $G_0$; this leads to the conformal invariance of the G-CAT
models, which generalizes the CAT models~\ct{AFGZ,bb} for non-abelian groups.

In section~\ref{sec:reduction}, we obtain the G-CAT models via Hamiltonian
reduction of the two-loop WZNW model, where the fields are elements of a
Kac-Moody group and the corresponding current algebra is a two-loop Kac-Moody
algebra~\ct{AFGZ,schw}. The Hamiltonian reduction can be
characterized by the choice of a gradation of the Kac-Moody algebra $\cgh$ and
of the elements
$\Lambda_{\pm l}$ appearing in~\rf{breakterm}; the reduction can be performed
for a general gradation, but we restrict ourselves to the case of integer
gradations. The constraints are imposed  on some non-chiral quantities which
have a simpler algebraic structure than the currents. Actually, the
generators $\Lambda_{\pm l}$ correspond to the constant value of those
quantities in the subspaces of grade $\pm l$. When $l$ is greater than the
lowest positive grade, the constraints on the currents are of second class.
However, not all the components of the currents with grade between zero and
$l$ are constrained, implying that the field $B$ is not the only degree of
freedom of the reduced model. The extra degrees of freedom correspond to two
sets of chiral fields, one set for each chirality, but these chiral fields
are decoupled from the field $B$.

In section~\ref{sec:confinv}, we show that the reduced model is
conformally invariant because the current associated to the grading
operator can be used to improve the stress tensor allowing the constraints to
weakly commute with it. We also show that the
field in the direction of the grading operator, called $\eta$,  is free;
even more, its interaction with the other fields is such that for every
regular solution of $\eta$ one can eliminate it from the remaining equations
of motion by a coordinate transformation. Then, the resulting model is not
conformally invariant and we refer to it as the Generalized non-abelian
Affine Toda (G-AT). All this is a generalization of what occurs in the
abelian Conformal Affine Toda models~\ct{con}.

The general solution of the G-CAT models is constructed in
section~\ref{sec:lssol} using the method of Leznov and
Saveliev~\ct{misha,misha2,misha3}. For $l$ not being the lowest positive grade,
the number of chiral parameters of the solution is not equal to twice the
dimension of $G_0$. There is a number of extra parameters equal to the number
of chiral fields appearing in the Hamiltonian reduction described above.  Those
parameters are present even though we start from a zero curvature connection
depending only on the field $B$; their origin has to do just with the algebraic
structure of the connections.

The symmetries of the G-CAT models are discussed in
section~\ref{sec:symmetries}; they correspond to the left (right)
translations on the Kac-Moody group which leave $\Lambda_{-l}$
($\Lambda_{l}$) invariant. One of our main interests is to construct the
one-soliton solutions of the model, which are static solutions in some
particular Lorentz frame. We show that the condition for the existence
of static solutions is that $\Lambda_{\pm l}$  can be transformed by some
constant element of $G_0$ into some elements $E_{\pm l}$  such that $\lb
E_{l} \, , \, E_{-l}\rb$ is in the direction of the central term only. This
condition is easily satisfied if $E_{\pm l}$ lie on a Heisenberg subalgebra
of $\cgh$.

The fundamental particles of the G-CAT model are obviously massless due to
conformal invariance. In section~\ref{sec:classmass} we show that the
classical masses of the fundamental particles of the G-AT models appear as a
consequence of the spontaneous breakdown of the conformal symmetry, resembling
very much the Higgs mechanism in gauge theories. We show that the masses are
proportional to the eigenvalues of the operator   $\lb E_{l} \, , \,\lb
E_{-l}\, , \,\cdot\rb\rb$ when acting on the Lie algebra of $G_0$; the
proportionality constant being the vacuum expectation value of the
(Higgs-like) field $e^{l\,\eta}$.

In section~\ref{sec:dressing} we use the dressing transformation
method~\ct{dress1,dress2,dress3,dress4} to construct the solutions of the
G-CAT  model lying in the orbit of the vacuum. The construction is quite simple
and leads to a very useful expression for the solutions.  In fact, it allows
us to show that the {\it solitonic specialization} of the Leznov-Saveliev
solution described in~\ct{otu,osu} leads to all solutions in the orbit of the
vacuum.

The results obtained with the dressing method guide us to the introduction of
the $\tau$-function~\ct{miwa,kw} in section~\ref{sec:tau}. We define the
$\tau$-functions for the G-CAT models as the orbit of the highest weight
state of an integrable representation of the Kac-Moody $\cgh$ obtained under
the action of the group element used in the dressing transformation. The
particular integrable representation is the one associated to the gradation
of $\cgh$ used in the Hamiltonian reduction, in a manner explained in the
appendix~\ref{ap:km}. One of the nice features of our definition is that it
is independent of the way the integrable representation is realized; in
particular, it is independent of the level of the representation. In this
sense, our results allow the generalization of the connection between
$\tau$-functions and zero-curvature integrable hierarchies of
partial differential equations worked out in~\ct{tault}.

The masses of the one-soliton solutions are
calculated in section~\ref{sec:solmass}. Following~\ct{acfgz}, we explore the
conformal invariance of the G-CAT model to show that the masses of the
solitons of the G-AT models come from a total divergence which is given
by the improvement term of the G-CAT stress tensor. We find an explicit
formula for those masses by using the highest weight state of an integrable
representation of $\cgh$ that is dual to the one used in the
$\tau$-function definition. Then, the one-soliton solutions can be put in one
to one correspondence with the massive fundamental particles. In addition,
non-vanishing soliton masses are proportional to the masses of their
corresponding fundamental particles. This is a generalization of
what occurs in the abelian Affine Toda models~\ct{acfgz,otu}, and suggests
some deep structure in such theories which is still not well understood;
they could be related to a {\it duality} transformation similar to the one
conjectured in four-dimensional non-abelian gauge theories~\ct{duality,watts}.

In section~\ref{sec:examples} we discuss the G-CAT models associated to the
principal and homogeneous gradations of any Kac-Moody algebra $\cgh$. In the
homogeneous case, the one-soliton solutions are studied in great
detail, also considering the particular case when $G_0$ is compact. Finally, we
present our conclusions in section~\ref{sec:discuss}.

In appendix~\ref{sec:twovir} we show how to use the structure of the two-loop
Kac-Moody current algebra to construct what we call the two-loop Virasoro
algebra. In order to do that, we have to impose periodic boundary conditions
on the currents, and the structure of such algebra is more complex when the
two central terms of the two-loop current algebra are incommensurable; the
role of this algebra in relation to the symmetries of the G-CAT model has
not been investigated yet. Finally, appendices~\ref{ap:km} and~\ref{ap:gcero}
include our conventions and some particular properties of Kac-Moody algebras
used along the paper.

\sect{Generalized  Reduction of Two-Loop WZNW model}
\label{sec:reduction}

\ \indent
The two-loop WZNW model was introduced in~\ct{AFGZ} as the
generalization of the ordinary WZNW model to the affine case. Its equations of
motion are given by
\be
\pa_{+} \( \pa_{-} {\hat g} \, {\hat g}^{-1} \) = 0 \qquad;\qquad
\pa_{-} \({\hat g}^{-1}  \pa_{+} {\hat g} \) = 0
\lab{motion}
\ee
where $\pa_{\pm}$ are derivatives with respect to the light-cone variables
$x_{\pm} = x\pm t$, and ${\hat g}$ is an element of the group $G$
formed by exponentiating an untwisted affine (real) Kac-Moody (KM) algebra
$\kmlie$. Its generators $T_{a}^m$, $D$, and $C$ satisfy the commutation
relations
\br
\lbrack T^m_a \, , \, T^n_b \rbrack &=& f_{ab}^c\> T_c^{m+n} + m\>C\, g_{ab}
\> \delta_{m+n,0} \lab{affina}\\
\lbrack D \, , \, T^m_a \rbrack &=& m\> T^m_a,\qquad \lbrack C \, , \, D
\rbrack \,=\, \lbrack C \, , \, T^m_a \rbrack = 0 \lab{affind}
\er
where $f^c_{ab}$ are the structure constants of a finite (real) semisimple Lie
algebra $\lie$, $n$ and $m$ are integers, and $g_{ab}$ is the Killing form of
$\lie$, {\it i.e.\/}, $g_{ab} = \Tr (T_a T_b)$, $T_a$ being the generators of
$\lie$. The non-degenerate bilinear form of $\cgh$ is defined as (see also
\rf{trace1}-\rf{trace3})
\br
&&\Tr\left( T_{a}^m T_{b}^n\right) = \delta_{m+n,0}\> \Tr (T_a T_b),\qquad
\Tr\left(C\>D\right)=1 \nonu\\
&&\Tr\left( C\> T_{a}^m\right) = \Tr\left( D\> T_{a}^m\right)=0,
\er
and we will use the same notation, $\Tr$, for both the Killing form of $\cg$
and the bilinear form of $\cgh$.

The two-loop WZNW model is invariant under  left and right translations
\be
{\hat g}(x_{+},x_{-}) \ra {\hat g}_L(x_{-}) \, {\hat g} (x_{+},x_{-})
\, \, \, \,\,\, ,\,\,\,\,\, \,
{\hat g}(x_{+},x_{-}) \ra  {\hat g} (x_{+},x_{-}) \, {\hat g}_R(x_{+})
\lab{symwznw}
\ee
The corresponding Noether
currents are the components of $\pa_{-} {\hat g} \, {\hat g}^{-1}$ and ${\hat
g}^{-1} \pa_{+} {\hat g}$, and they generate two commuting copies of the so
called two-loop Kac-Moody algebra \ct{AFGZ}, defined by the relations
\br
\lbrack J^m_a (x) \, , \, J^n_b (y) \rbrack &=& f_{ab}^c\> J_c^{m+n} (x) \delta
(x - y)  +  g_{ab}\> \delta_{m,-n}\( k \pa_x  \delta (x - y)
+  m J^{{\cal C}}(x) \delta (x - y)\)
\phantom{......}\lab{kma}
\\
\lbrack J^{{\cal D}} (x) \, , \, J^m_a (y) \rbrack &=& m\> J^m_a (y)
\delta (x -y)
\lab{kmb} \\
\lbrack J^{{\cal C}} (x) \, , \, J^{{\cal D}} (y) \rbrack &=& k\> \pa_x
\delta (x - y)
\lab{kmc} \\
\lbrack J^{{\cal C}} ( x) \, , \, J^m_a (y) \rbrack &=& 0 \lab{kmd}
\er
The left and right currents satisfying the above relations are related to the
group element $\hat g$ in eq.\rf{motion} by
\br
{\cal J}_R(x_+) = k {\hat g}^{-1} \pa_{+} {\hat g} &=& \sum_{a,b } \sum_{n=-
\infty}^{\infty} g^{ab} J^{-n}_{R , a}(x_{+}) T^n_{b}  +
J^{{\cal D}}_{R} (x_{+}) C + J^{{\cal C}}_{R}(x_{+}) D\lab{currplus} \\
{\cal J}_L(x_-) = -k \pa_{-} {\hat g}\, {\hat g}^{-1}  &=& \sum_{a,b }
\sum_{n=- \infty}^{\infty} g^{ab} J^{-n}_{L , a}(x_{-}) T^n_{b}  +
J^{{\cal D}}_{L}(x_{-}) C + J^{{\cal C}}_{L}(x_{-}) D \phantom{....}
\lab{currminus}
\er
where $g^{ab}$ is the inverse of the Killing form $g_{ab}$ defined above. The
different meaning of the two central extensions in eqs.\rf{kma}-\rf{kmd}
algebra is clarified by expressing the algebra as
\be
\Bigl[ \Tr\left(U{\cal J}(x)\right)\,,\, \Tr\left(V{\cal J}(y)\right)\Bigr]
\,=\, \Tr\Bigl(\lbrack U, V\rbrack\, {\cal J}(x)\Bigr)\, \delta(x-y) \,+\,
k\, \Tr\left(UV\right)\, \partial_x\delta(x-y),
\lab{otheralg}
\ee
where $U,V$ are two elements of the Kac-Moody algebra $\cgh$,  $\cal J$ is
either ${\cal J}_R$ or ${\cal J}_L$, and $\Tr$ is the invariant bilinear
form of $\cgh$.

Consider now a gradation of the Kac-Moody algebra $\kmlie$
\be
\kmlie = \bigoplus_{s}  \,  \kmlie_s
\lab{grade}
\ee
with
\be
\lb \kmlie_s \, , \, \kmlie_r \rb \subset  \kmlie_{s+r}
\lab{grade2}
\ee
The reduction presented in this section does not require that this
gradation is integer; it just needs that the grades $s$ take zero,
positive and negative values, {\it i.e.\/},
\be
\cgh = \cgh_{+} \oplus \cgh_{0} \oplus \cgh_{-}
\lab{grade3}
\ee
with
\be
 \kmlie_+ \equiv \bigoplus_{s>0} \kmlie_s \, , \qquad
\kmlie_- \equiv \bigoplus_{s<0} \kmlie_s
\lab{grade4}
\ee
Nevertheless, in the rest of the paper, we shall restrict ourselves to
integer gradations, which can be described in a very systematic way using
the results of~\ct{kac1,kac2} (see appendix~\ref{ap:km}).

We now consider those group elements that can be written  in a ``Gauss
decomposition'' form
\be
{\hat g} = NBM \in G
\lab{gauss}
\ee
where $N$, $B$ and $M$ are group elements formed by exponentiating elements
of $\kmlie_+$, $\kmlie_0$ and $\kmlie_-$  respectively. It is well known that
the use of the Gauss decomposition requires that the group has definite
properties regarding its compactness. Nevertheless, when the subspaces of
$\cgh$
with opposite grades are non-degenerately paired by the bilinear form, as it
happens when the gradation is integer, the validity of the Gauss decomposition
does not imply any condition regarding the compactness of
the subgroup formed by exponentiating the elements of
$\cgh_0$, $G_0$; it only requires that the restriction of the bilinear form to
$\cgh_+\oplus \cgh_-$ is maximally non-compact. When the gradation is integer
(see~\rf{grad1}-\rf{grad2}), this means that $\{H_{1}^n,\ldots,H_{r}^n\}$, for
all $n\not=0$, and
$\{E_{\alpha}^{m}\}$, with $m\>N_{\bf s} + \alpha\cdot H_{\bf s}\not=0$,
must be among the generators of the chosen real form of the algebra $\cgh$.
Then, one is free to choose the subgroup $G_0$ to be either compact or
non-compact; each choice leading to different theories. In
section~\ref{sec:symmetries}, we shall come back to this point, which is very
important in relation to the physical interpretation of the resulting
models~\ct{tlp}.

Using eq.\rf{gauss}, we can write the equations of motion \rf{motion} as
\br
\pa_- K_R &=& - \lb K_R \, , \, \pa_- M M^{-1} \rb
\lab{kreq}\\
\pa_+ K_L &=&  \lb K_L \, , \,  N^{-1}\pa_+ N \rb
\lab{kleq}
\er
where we have introduced
\br
K_L &\equiv& N^{-1} \pa_{-} {\hat g} \, {\hat g}^{-1} N \nonu\\
&=& N^{-1}\pa_- N + \pa_- B B^{-1} + B \pa_- M M^{-1} B^{-1}
\lab{kl}\\
K_R &\equiv&  M {\hat g}^{-1} \pa_{+} {\hat g}M^{-1} \nonu\\
&=& B^{-1} N^{-1}\pa_+ N B + B^{-1}\pa_+ B  + \pa_+ M M^{-1}
\lab{kr}
\er
Although the quantities $K_{L/R}$ are not chiral, they have a simpler
structure than the currents and will be very useful in what follows. We will
reduce the two-loop WZNW model by imposing constraints not directly on the
currents but on $K_{L/R}$. We impose the constraints
\br
B^{-1} \( N^{-1}\pa_+ N\) B &=& \Lambda_l
\lab{const1}\\
B \(\pa_- M\) M^{-1} B^{-1} &=& \Lambda_{-l}
\lab{const2}
\er
where $\Lambda_{\pm l}$ are constant elements of $\kmlie_{\pm l}$ ($l>
0$). These constraints reduce the two-loop WZNW model to a theory containing
only the fields corresponding to the components of $B$ and to the components
of $N$ and $M$ associated to the generators whose grade is $< l$ and $>-l$
respectively.

To obtain the equations of motion for such model one notices that
the constraints \rf{const1} and \rf{const2} imply that
\br
 N^{-1}\pa_+ N &\in& \cgh_l
\lab{const3}\\
 \(\pa_- M\) M^{-1} &\in& \cgh_{-l}
\lab{const4}
\er
Therefore the only terms of zero grade on the right hand side of \rf{kleq} are
coming from  $\lb \Lambda_{-l} \, ,\, N^{-1}\pa_+ N \rb \, = \, \lb
\Lambda_{-l} \, ,\, B \Lambda_l B^{-1} \rb$. So, we get
\be
\pa_{+}\( \pa_{-} B B^{-1}\) = \lb \Lambda_{-l} \, , \, B \Lambda_l B^{-1}\rb
\lab{natoda1}
\ee
which can also be written as
\be
\pa_{-}\(B^{-1} \pa_{+} B \) = -\lb \Lambda_{l} \, , \, B^{-1} \Lambda_{-l} B
\rb
\lab{natoda2}
\ee
These are the equations of motion of what we call the {\em generalized
non-abelian conformal affine Toda models} (G-CAT). When the gradation is
integral and $l=1$, these equations were first considered in~\ct{misha,misha2},
but not including the field in the direction of the derivation
$D$, which leads  to the conformal invariance of the model as we explain in the
next section. In addition, the lagrangian interpretation of these
equations was not studied in those references either.

As for the equations of motion of the fields corresponding to $N$ and $M$, we
obtain from
\rf{kreq} and \rf{kleq} that
\be
\pa_{-} \( \pa_{+} M M^{-1} \)_{>-l} = 0 \qquad
\pa_{+} \( N^{-1} \pa_{-} N  \)_{<l} = 0
\ee
However, one can get more information on the dynamics of such fields from the
constraints \rf{const1} and \rf{const2}. We write
\br
N \equiv \exp (\sum_{s>0} \zeta_s) \,\,\,\, &,& \,\,\,\,\zeta_s \in \kmlie_s
\nonu\\
M \equiv \exp (\sum_{s>0} \xi_{-s}) \,\,\,\, &,& \,\,\,\, \xi_{-s} \in
\kmlie_{-s}
\er
and we label the negative grades in a ordered way as $-s_1>-s_2>-s_3>\cdots$.
Using the fact that $\pa e^{T} e^{-T}= \pa T + {1\o 2!}\lb T\, ,\,\pa T \rb +
{1\o 3!}
\lb T \lb T\, , \, \pa T \rb \rb + \cdots$,
one observes that the term in $\pa_- M M^{-1}$ whose grade is maximal is
just $\pa_{-} \xi_{-s_{1}}$. Therefore, if $s_1< l$, \rf{const4} requires
that $\pa_{-}\xi_{-s_{1}}=0$. Using this result, the term whose grade is next
to $-s_1$ in
$\pa_- M M^{-1}$ is  $\pa_{-}\xi_{-s_{2}}$ and, therefore, if $s_2<l$,
\rf{const4} implies $\pa_{-}\xi_{-s_{2}}=0$. So, following such reasoning, one
gets
\be
\pa_{-}\xi_{-s}=0 \qquad \qquad \mbox{\rm for $s<l$}
\lab{xichiral}
\ee
Analogously, for the fields in $N$ one gets
\be
\pa_{+}\zeta_{s}=0 \qquad \qquad \mbox{\rm for $s<l$}
\lab{zetachiral}
\ee
Therefore the extra fields in $N$ and $M$ are chiral and decouple from
the $B$ fields. Obviously, when $l$ corresponds to the lowest positive grade,
the reduced model contains only $B$ and such chiral fields are absent.

{}From \rf{kl}-\rf{kr} and \rf{const1}-\rf{const2} one obtains
\br
\pa_{-} {\hat g} \, {\hat g}^{-1} &=&
 N \(  \Lambda_{-l}  + \pa_- B B^{-1}  \) N^{-1} + \pa_- N N^{-1} \nonu\\
&\equiv& \Lambda_{-l} + \sum_{s>-l} j_L^{(s)}
\lab{lcurr}\\
{\hat g}^{-1} \pa_{+} {\hat g} &=&
M^{-1}\( \Lambda_{l}  + B^{-1}\pa_+ B \) M + M^{-1}\pa_+ M   \nonu\\
&\equiv& \Lambda_{l} + \sum_{s<l} j_R^{(s)}
\lab{rcurr}
\er
where the index $s$ denotes the grade of the current component.

Therefore, in terms of the currents, the constraints \rf{const1}-\rf{const2}
are
\br
\(\pa_{-} {\hat g} \, {\hat g}^{-1}\)_{-l<s<0} &=& \( N   \Lambda_{-l}
 N^{-1}\)_{-l<s<0}
\lab{curconst1a}\\
\(\pa_{-} {\hat g} \, {\hat g}^{-1}\)_{-l} &=& \Lambda_{-l}
\lab{curconst1b}\\
\(\pa_{-} {\hat g} \, {\hat g}^{-1}\)_{<-l} &=& 0
\lab{curconst1c}
\er
and
\br
\({\hat g}^{-1} \pa_{+} {\hat g}\)_{>l} &=&  0
\lab{curconst2a}\\
\({\hat g}^{-1} \pa_{+} {\hat g}\)_{l} &=& \Lambda_{l}
\lab{curconst2b}\\
\({\hat g}^{-1} \pa_{+} {\hat g}\)_{0<s<l} &=& \( M^{-1} \Lambda_{l}
 M \)_{0<s<l}
\lab{curconst2c}
\er
When $l$ is not the lowest positive grade in \rf{grade}, these constraints
are not only first class, but they also include second class constraints.

\sect{Conformal Invariance of the G-CAT models}
\label{sec:confinv}

\ \indent
The two-loop WZNW model \rf{motion} is a conformally invariant theory and
the corresponding Virasoro generators are given by the Sugawara construction.
The procedure to show that the G-CAT models are also conformal invariant is the
same for both chiralities, and, here, we consider only the right moving
component of the stress tensor. From now on, we shall restrict ourselves to the
models defined in terms of integral gradations (see appendices~B and~C).

Using the Sugawara construction we can obtain two Virasoro generators from the
currents of the two-loop Kac-Moody algebra (see \rf{kma}-\rf{kmd}
and \rf{currplus})
\br
T^{(1)} (x) &=& {1 \o 2 k}  \sum_{a,b=1}^{{\rm dim}\, \lie}
\sum_{n=- \infty}^{\infty} g^{ab} J^n_{R,a} (x) J^{-n}_{R,b} (x)
\lab{vir1}\\
T^{(2)} (x) &=& {1 \o  k} J^{{\cal D}}_R(x) J^{{\cal C}}_R (x)
\lab{vir2}
\er
They both satisfy a centreless Virasoro algebra
\be
\lbrack T^{(i)} (x) \, , \, T^{(i)} (y) \rbrack = 2\> T^{(i)} (y) \pa_x \delta
(x - y) -  \pa_y \(T^{(i)} (y)\) \delta (x - y) \, ;
\qquad \mbox{\rm for $i=1,2$}
\lab{virai}
\ee
and they commute
\be
\lbrack T^{(1)} (x) \, , \, T^{(2)} (y) \rbrack = 0
\ee
These Virasoro generators induce the following transformation of the currents:
\br
\lbrack T^{(1)} (x) \, , \, J^n_{R,a} (y) \rbrack &=&
 J^n_{R,a} (y) \d^{\pr} (x-y) - \pa_y \( J^n_{R,a} (y) \) \d (x-y)
\nonu\\
&-& {n\o k} J^n_{R,a} (y)  J^{{\cal C}}_R (y) \d (x-y)\\
\lbrack T^{(1)} (x) \, , \, J^{{\cal D,C}}_R (y) \rbrack &=& 0
\er
and
\br
\lbrack T^{(2)} (x) \, , \, J^n_{R,a} (y) \rbrack &=&
{n\o k} J^n_{R,a} (y)  J^{{\cal C}}_R (y) \d (x-y) \nonu\\
\lbrack T^{(2)} (x) \, , \, J^{{\cal D,C}}_R (y) \rbrack &=&
J^{{\cal D,C}}_R (y) \d^{\pr} (x-y) - \pa_y \( J^{{\cal D,C}}_R (y)\) \d (x-y)
\er

Therefore all the currents transform as primary fields of conformal weight $1$,
under the sum of the two Virasoro generators
\be
T (x) \equiv  T^{(1)} (x) +  T^{(2)} (x)
\lab{suga}
\ee
and so
\be
\lbrack T (x) \, , \,  {\cal J} (y) \rbrack =  {\cal J} (y) \pa_{x} \delta
(x - y) -  \pa_y \( {\cal J} (y)\) \delta (x - y)   \lab{confa}
\ee
where ${\cal J} (y)$ stands for any of the currents. Obviously $T(x)$ also
satisfies a centreless Virasoro algebra
\be
\lbrack T (x) \, , \, T (y) \rbrack = 2 T (y) \pa_x \delta
(x - y) -  \pa_y \( T (y)\) \delta (x - y)
\lab{vira}
\ee

In \rf{curconst2b} the currents in the direction of the generators of grade $l$
were set to a constant. This breaks the conformal invariance associated to
$T(x)$ since such currents are not scalars. However, we can modify $T(x)$ to
obtain a new Virasoro generator under which those currents are scalars.
If the gradation \rf{grade} is realized by a grading operator $Q_{\bf s}$,
we can use the  component of the current corresponding to $Q_{\bf s}$,
{\it i.e.\/},
\be
J_{R}^{Q_{\bf s}}(x) \,= \, \Tr\left( Q_{\bf s}{\cal J}_R (x)\right),
\ee
to modify $T(x)$; then, from \rf{currplus} and \rf{otheralg}, the right
currents
in the direction of the generators of grade $l$ have grade $-l$ with respect to
$J_{R}^{Q_{\bf s}}$
\be
\lbrack J_{R}^{Q_{\bf s}} (x) \, , \,  J_R^{(-l)} (y) \rbrack = -l \,
J_R^{(-l)} (y)
\delta (x - y).
\lab{gradel}
\ee
Therefore, we introduce the improved stress tensor as
\be
L(x)\, \equiv\, T(x)\, +\, {1\o l} \pa_x J_{R}^{Q_{\bf s}} (x)
\lab{imprvira}
\ee
under which the currents set to constant are scalars.

The improved stress tensor \rf{imprvira} satisfies
\be
\lbrack L (x) \, , \, L (y) \rbrack = 2 L (y) \pa_x \delta
(x - y) -  \pa_y L (y) \delta (x - y),
\lab{viraimpr}
\ee
which is a centreless Virasoro algebra too; notice that the Virasoro algebra
generated by $L(x)$ might have a central extension proportional to
$\Tr(Q_{\bf s}^2)$, but, for the particular choice \rf{choicegrad}, it
vanishes. With respect to $L(x)$, the components of the current
${\cal J}_R(x)$ whose grade is $j\in \IZ$ transform as primary fields of
conformal weight $1-j/l$, with the exception of the component $\Tr\left(C{\cal
J}_R(x)\right)$ whose transformation is
\br
\Bigl[ L(x), \Tr\left(C{\cal J}_R(y)\right) \Bigr]\, &=& \,
\Tr\left(C{\cal J}_R(y)\right) \partial_x\delta(x-y) \nonu \\
&&-\, \partial_y \biggl(\Tr\left(C{\cal J}_R(y)\right)\biggr) \delta(x-y)
+\,{kN_{\bf s}\over l}\, \partial_{x}^2\delta(x-y).
\lab{anomalous}
\er
For the other chirality a similar procedure
applies, thus establishing the conformal invariance of the G-CAT models.

According to eq.\rf{paramgroup}, we parameterize the field $B$ as
\be
B \equiv B_0 \, e^{\nu\> C\, +\, \eta\> Q_{\bf s}}
\lab{parb}
\ee
where $B_0$ denotes an element of the subgroup formed by exponentiating the
elements of the subalgebra $\cgh_{0}^{\ast}$ defined in
appendix~\ref{ap:gcero}. From
\rf{natoda1}, we get
\be
\pa_{+} \( \pa_{-} B_0 B_0^{-1} \) + \pa_{+} \pa_{-}\nu\, C
+ \pa_{+} \pa_{-}\eta \, Q_{\bf s}   = e^{l
\eta}
\,\lb \Lambda_{-l} \, , \, B_0 \Lambda_{l} B_0^{-1} \rb
\lab{eqb}
\ee
The conformal invariance of the G-CAT model can be made  explicit by verifying
that the above equation is invariant under the conformal transformations
\be
x_{+} \ra {\tilde x}_{+} = f(x_{+}) \, ; \qquad x_{-} \ra {\tilde x}_{-} =
g(x_{-})
\ee
if the fields transform as
\br
B_0(x_{+},x_{-}) &\ra& {\tilde B_0}({\tilde x_{+}},{\tilde x_{-}}) =
B_0(x_{+},x_{-})\nonu\\
e^{-\nu(x_{+},x_{-})} &\ra& e^{-{\tilde \nu}({\tilde x_{+}},{\tilde x_{-}})} =
\( f^{\pr}(x_{+})\)^{\d} \( g^{\pr}(x_{-})\)^{{\bar \d}}
e^{-\nu(x_{+},x_{-})}\nonu\\
e^{-\eta (x_{+},x_{-})} &\ra& e^{-{\tilde \eta}
({\tilde x}_{+},{\tilde x}_{-})} = \( f^{\pr}(x_{+})\)^{1/l}
\( g^{\pr}(x_{-})\)^{1/l} e^{-\eta (x_{+},x_{-})}
\lab{primary}
\er
where $\d$ and ${\bar \d}$ are arbitrary.
Notice that $\Tr\left(C{\cal J}_{R/L}(x_\pm)\right) =
k\> N_{\bf s}\>\partial_{\pm} \eta(x_+,x_-)$, and that the transformation
expressed by eq.\rf{primary} is in agreement with eq.\rf{anomalous}.

The grading operator $\qs$ has a component in the direction of $D$ that
cannot be the result of any commutator, {\it i.e.\/}, $D, \qs \notin
[\cgh,\cgh]$; then, it follows from \rf{eqb} that $\eta$ is actually a free
field
\be
\pa_{+} \pa_{-}\eta =0
\lab{eta}
\ee
Therefore the solutions for $\eta$ are of the form
\be
\eta (x_{+},x_{-}) = \eta_{+}(x_{+}) + \eta_{-}(x_{-}),
\lab{etasol}
\ee
and, for every regular solution, the term $e^{l\eta}$ can be eliminated by a
coordinate transformation, like in the  usual CAT models~\ct{con}. In fact, for
regular solutions, we can define the coordinate transformation
\be
x_{+} \ra {\tilde x}_{+} \equiv\int^{x_{+}} d y_{+} e^{l\,\eta_{+}} \qquad
\qquad  x_{-} \ra {\tilde x}_{-} \equiv\int^{x_{-}} d y_{-} e^{l\,\eta_{-}}
\lab{transf}
\ee
and so
\be
\pa_{+}\( A \pa_{-}B\) \ra e^{l\, (\eta_{+} + \eta_{-})}
{\tilde \pa}_{+}\( A{\tilde \pa}_{-}B\)
\ee
Hence, when the equation of motion for $\eta$ holds, \rf{eqb} becomes
\be
{\tilde \pa}_{+} \( {\tilde \pa}_{-} B_0 B_0^{-1} \)\, +\, {\tilde
\pa}_{+}{\tilde \pa}_{-} \nu\> C    =  \,\lb \Lambda_{-l} \, ,\, B_0
\Lambda_{l}
B_0^{-1} \rb
\lab{gat}
\ee
We will refer to such theory, containing only the fields $B_0$ and $\nu$, as
the
{\em Generalized non-abelian affine Toda models} (G-AT). Let us point out
that, in \rf{gat}, the dynamics of the field $B_0$ is actually independent of
the field $\nu$, while the equation of motion of $\nu$ is, see
\rf{paramalg}-\rf{projection},
\be
{\tilde\pa}_+{\tilde\pa}_- \nu = {1\over N_{\bf s}} \Tr \left(\qs\lb
\Lambda_{-l} \, ,\, B_0 \Lambda_{l} B_0^{-1} \rb \right) = -{l\over N_{\bf s}}
\Tr \left(\Lambda_l B_{0}^{-1} \Lambda_{-l} B_0 \right).
\lab{motionnu}
\ee
It is also worth noticing that the G-AT model is already invariant under the
scale transformation $x_{\pm}\ra \lambda^{\pm1} x_\pm$ (which is,
in fact, a Lorentz transformation), and that the field $\eta$ actually plays
the role of a connection to implement conformal invariance.

\sect{The Leznov-Saveliev solution}
\label{sec:lssol}

\ \indent
In this section, we use the method of Leznov and Saveliev~\ct{misha2,misha3} to
obtain the general solution for the G-CAT models. This method has been
discussed quite extensively in the context of Toda models and we present it
here in some detail because, when applied to the G-CAT model, it has some
features not common to the other ones. Specifically, when $l$, the grade of the
currents set to a constant value (see
\rf{curconst1b}-\rf{curconst2b}), is not the lowest non-vanishing grade, the
solution  has a number of  chiral parameters greater than twice the number of
fields. In addition, we will use the results of this section to relate some
special Leznov-Saveliev solutions to the dressing transformations and
$\tau$-functions discussed in sections~\ref{sec:dressing} and \ref{sec:tau}.

The equations of motion \rf{natoda1} (or equivalently \rf{natoda2}) can be
put in the form of a zero-curvature condition
\be
\lb \pa_{+} + A_{+} \, , \, \pa_{-} + A_{-} \rb = 0
\lab{zc}
\ee
with the gauge potentials being given by
\br
A_{+} &=& - B \Lambda_l B^{-1} \nonu\\
A_{-} &=& - \pa_{-}B B^{-1} + \Lambda_{-l}
\lab{gp}
\er
Notice that the zero-curvature condition \rf{zc} is equivalent to the
equations of motion for the $B$ fields only. It does not involve the equations
for the chiral fields $\xi_{-s}$ and $\zeta_{s}$  ($1<s<l$) given by
\rf{xichiral}-\rf{zetachiral}.

Since the gauge potentials \rf{gp}  have to satisfy \rf{zc}, they must be
of the ``pure gauge'' form, {\it i.e.\/},
\be
A_{\pm} = - \pa_{\pm} g g^{-1}
\lab{pg}
\ee
where $g$ is an exponentiation of the Kac-Moody algebra generators
\rf{affina}-\rf{affind}. The element $g$ is obtained by integrating \rf{pg} a
la Dyson, {\it i.e.\/}, $g(x_{+},x_{-}) = g(0) P \exp \( \int^{(x_{+},x_{-})}
dy^{\mu} A_{\mu}\)$. However, since the field strength associated to $A_{\mu}$
vanishes,  it follows from Stoke's theorem that the integral is path
independent. In general, due to the form of $A_{\mu}$ expressed by eq.\rf{gp},
the integration becomes easier by choosing two special paths. Here, we use an
algebraic procedure to integrate \rf{pg}, and we start by writing $g$ in terms
of two different group elements
$g_1,g_2\in G$ as
\be
g \equiv g_1 \equiv B g_2,
\lab{2path}
\ee
which are related to the two previously mentioned special paths. Then, we get
\br
\pa_{\pm} g g^{-1} &=& \pa_{\pm} g_1 g_1^{-1}\nonu\\
\pa_{\pm} g g^{-1} &=& \pa_{\pm}B B^{-1} + B\pa_{\pm} g_2 g_2^{-1}B^{-1}
\er
Therefore from \rf{gp} and \rf{pg}
\br
\pa_{+} g_1 g_1^{-1} &=& B \Lambda_l B^{-1}\lab{comp1}\\
\pa_{-} g_1 g_1^{-1} &=& \pa_{-}B B^{-1} - \Lambda_{-l}\lab{comp2}\\
\pa_{+} g_2 g_2^{-1} &=& -B^{-1}\pa_{+}B  + \Lambda_{l}\lab{comp3}\\
\pa_{-} g_2 g_2^{-1} &=& -  B^{-1}\Lambda_{-l} B \lab{comp4}
\er

We now consider the states of a highest weight representation of the
Kac-Moody algebra $\cgh$ which are annihilated by the positive grade
generators, {\it i.e.\/},
\be
T \mid \mu \rangle = 0 \, ; \qquad \qquad \mbox{\rm for $T \in \cgh_{+}$}
\lab{hws}
\ee
In addition, we require that
\be
\langle  \mu \mid T = 0 \, ; \qquad \qquad \mbox{\rm for $T \in \cgh_{-}$}
\lab{lws}
\ee
which is a consequence of \rf{hws} if $\cgh_{-}$ is related to $\cgh_{+}$ by
conjugation. The set of such highest weight states constitutes a representation
of the finite subalgebra $\cgh_0$, since if $\mid \mu \rangle$ satisfies
\rf{hws} so does $\cgh_0 \mid \mu \rangle$; we will assume that this
representation of $\cgh_0$ is faithful, which is not true for the integrable
highest weight representations mentioned in appendix~\ref{ap:km}.

{}From \rf{comp1}, \rf{comp4}, \rf{hws} and \rf{lws}, one then gets
\br
\pa_{+} g_1 g_1^{-1}\mid \mu \rangle &=& - g_1\pa_{+} g_1^{-1}\mid \mu \rangle
= B \Lambda_l B^{-1} \mid \mu \rangle = 0 \nonu\\
\langle \mu \mid \pa_{-} g_2 g_2^{-1} &=& - \langle \mu \mid
B^{-1}\Lambda_{-l} B
= 0
\er
and so the states $g_1^{-1}\mid \mu \rangle$ and $\langle \mu \mid  g_2$ are
chiral
\be
\pa_{+} g_1^{-1}\mid \mu \rangle = 0 \, ; \qquad \qquad
\langle \mu \mid \pa_{-} g_2 = 0
\ee
Now, from \rf{2path},
\be
\langle \mu^{\pr} \mid B^{-1}\mid \mu \rangle = \langle \mu^{\pr} \mid g_2
g_1^{-1}\mid
\mu \rangle
\lab{matrixb}
\ee
and so, the matrix elements of $B^{-1}$ are obtained by contraction of two
chiral vectors.

Next, let us use the Gauss decomposition to write
\be
g_1 = \cn \, B_{-} M_{-} \, ; \qquad g_2 = \cm B_{+} N_{+}
\lab{gauss12}
\ee
where
\be
B_{\pm} \in \exp \left(\cgh_0\right), \qquad
\cn \, , N_{+} \in \exp \left(\cgh_{+}\right), \qquad
\cm \, , M_{-} \in \exp \left(\cgh_{-}\right)
\ee
Substituting \rf{gauss12} into \rf{comp1}, one gets
\be
\cn^{-1} B \Lambda_l B^{-1} \cn = \cn^{-1} \pa_{+} \cn + \pa_{+} B_{-}
B_{-}^{-1} + B_{-} \pa_{+} M_{-} M_{-}^{-1}B_{-}^{-1}
\ee
and, consequently,
\br
\pa_{+} B_{-} B_{-}^{-1} &=& 0\lab{comp1a}\\
\pa_{+} M_{-} M_{-}^{-1} &=& 0\lab{comp1b}\\
\pa_{+} \cn \cn^{-1} &=& B \Lambda_l B^{-1}
\lab{comp1c}
\er
Analogously, substituting \rf{gauss12} into \rf{comp4} one gets
\be
-\cm^{-1}  B^{-1}\Lambda_{-l} B  \cm = \cm^{-1} \pa_{-} \cm + \pa_{-} B_{+}
B_{+}^{-1} + B_{+} \pa_{-} N_{+} N_{+}^{-1}B_{+}^{-1}
\ee
and, so,
\br
\pa_{-} B_{+} B_{+}^{-1} &=& 0\lab{comp4a}\\
\pa_{-} N_{+} N_{+}^{-1} &=& 0\lab{comp4b}\\
\pa_{-} \cm \cm^{-1} &=& -  B^{-1}\Lambda_{-l} B \lab{comp4c}
\er

Now, from \rf{comp2},
\be
\cn^{-1} \( \pa_{-} B B^{-1} - \Lambda_{-l} \) \cn =
\cn^{-1} \pa_{-} \cn + \pa_{-} B_{-}
B_{-}^{-1} + B_{-} \pa_{-} M_{-} M_{-}^{-1}B_{-}^{-1}
\ee
therefore,
\be
\pa_{-} M_{-} M_{-}^{-1} = - B_{-}^{-1}\( \cn^{-1}\Lambda_{-l}\cn \)_{<0} B_{-}
\lab{eqform}
\ee
and, from \rf{comp3},
\be
\cm^{-1} \( - B^{-1} \pa_{+} B  + \Lambda_{l} \) \cm =
\cm^{-1} \pa_{+} \cm + \pa_{+} B_{+}
B_{+}^{-1} + B_{+} \pa_{+} N_{+} N_{+}^{-1}B_{+}^{-1}
\ee
which implies
\be
\pa_{+} N_{+} N_{+}^{-1} =  B_{+}^{-1}\( \cm^{-1}\Lambda_{l}\cm \)_{>0} B_{+}
\lab{eqforn}
\ee

Finally, if we write
\be
\cn = \exp \( \sum_{s>0} \chi^{-}_s \) \, ; \qquad
\cm = \exp \( \sum_{s>0} \chi^{+}_{-s} \)
\ee
with $\chi^{\pm}_{s} \in \cgh_{s}$,
we conclude from \rf{comp1c} and  \rf{comp4c} that
\br
\pa_{+} \chi^{-}_s &=& 0 \, ; \qquad \mbox{\rm for $0<s<l$}\nonu\\
\pa_{-} \chi^{+}_{-s} &=& 0 \, ; \qquad \mbox{\rm for $0<s<l$},
\er
by using arguments similar to those leading to \rf{xichiral} and
\rf{zetachiral}; these chiral quantities are the only parameters of $\cn$
and $\cm$  contributing in \rf{eqform} and \rf{eqforn} respectively.

Therefore, from \rf{matrixb} and the considerations above we conclude that
the general solution for the G-CAT model is given by
\be
\langle \mu^{\pr} \mid B^{-1}(x_{+},x_{-})\mid \mu \rangle = \langle \mu^{\pr}
\mid B_{+}(x_{+}) N_{+}(x_{+}) M_{-}^{-1}(x_{-}) B_{-}^{-1}(x_{-})\mid
\mu \rangle
\lab{solution}
\ee
where the quantities $N_{+}(x_{+})$ and $M_{-}(x_{-})$ are determined from
\rf{eqforn} and \rf{eqform} respectively. The group elements $B_{+}(x_{+})$
and $B_{-}(x_{-})$ together with the chiral quantities $\chi^{-}_s(x_{-})$ and
$\chi^{+}_{-s}(x_{+})$ for $0<s<l$, parameterize the general solution
\rf{solution}. When $l$ is the lowest positive grade of $\cgh$ in the
gradation, the parameters $\chi^{\pm}_s$ are absent.

Notice that, in order to reconstruct the group element $B$ in \rf{solution},
one has to choose a representation and a suitable number of states $\mid \mu
\rangle$. A priori, the above method does not give any criterium for that
choice.  In section \ref{sec:tau} we will see that, in the $\tau$-function
formalism, one needs just the highest weight state of a particular integrable
representation of the Kac-Moody algebra $\cgh$.

\sect{The symmetries and vacua of the G-CAT models}
\label{sec:symmetries}

\ \indent
The equations of motion \rf{natoda1}, or equivalently \rf{natoda2}, can be
obtained from the action
\br
S&=& {\kappa} \Biggl(-{1\over 8}\int d^2 \, x\> \Tr \( \pa_{\mu} B \pa^{\mu}
B^{-1}\) +  {1\o 12} \int d^3 \, y\> \epsilon^{ijk} \Tr \( B^{-1}\pa_{i} B\>
B^{-1}\pa_{j} B\> B^{-1}\pa_{k} B\) \Biggr.\nonu\\
&+&\Biggl. \int d^2 \, x\> \Tr \( \Lambda_{l}B^{-1} \Lambda_{-l}B\) \Biggr),
\lab{brokenwznw}
\er
{\it i.e.\/}, the WZNW action for the field $B\in \exp\left(\cgh_0\right)$
plus a potential, where, since $\Lambda_{\pm l}\notin \cgh_0$, $\Tr$ has
to be the trace form of the Kac-Moody algebra $\kmlie$. Nevertheless, the
calculation of almost all the terms in
\rf{brokenwznw} requires just the restriction of the trace form of $\cgh$ to
$\cgh_0$. Actually, the first two terms of
\rf{brokenwznw} involve generators of $\cgh_0$ only, and, in the last term,
the trace form can be evaluated on the subalgebra $\cgh_0$ except for a
constant term. Indeed, writing $B=e^T$ with $T\in \cgh_0$, one has
$\Tr \( \Lambda_{l}B^{-1} \Lambda_{-l}B\) = \Tr \( \Lambda_{l} \Lambda_{-l}\)
- \Tr \( \lb \Lambda_{-l}\, , \, \Lambda_{l}\rb T\) +
\h \Tr \( \lb\lb T \, , \,\Lambda_{-l}\rb \, , \,  \Lambda_{l}\rb T\) +
\cdots$.
In the action~\rf{brokenwznw}, notice that $\kappa$ is a coupling constant
which plays no role in the classical theory, where it can be considered just as
an overall factor.

The relation of the action~\rf{brokenwznw} with the affine Toda models can be
exhibited by using the parameterization~\rf{paramgroup} such that
\be
B= B_0 e^{\nu\> C\> + \> \eta\> Q_{\bf s}}, \qquad B_0 \in
\exp\left({\cgh_{0}^{\ast}}\right);
\ee
so, using eqs.\rf{paramWZNW1}-\rf{paramWZNW3},
\br
S&=& \kappa\Biggl(-{1\over8} \int d^2 \, x\> \Tr \( \pa_{\mu} B_0 \pa^{\mu}
B^{-1}_{0}\)
+  {1\o 12} \int d^3 \, y\> \epsilon^{ijk} \Tr \( B_0^{-1}\pa_{i} B_0\>
B^{-1}_{0}\pa_{j} B_0\> B^{-1}_{0}\pa_{k} B_0\) \Biggr.\nonu\\
&+& \Biggl. \int d^2\, x\>
e^{l\>\eta}\>\Tr \(\Lambda_{l}B^{-1}_{0}\Lambda_{-l}B_{0}\)\,
+ \,{1\over4} \>N_{\bf s}\>\int d^2 \, x\> \pa_{\mu}\>\nu \,\pa^{\mu}
\>\eta\Biggr).
\lab{brokenwznwplus}
\er
Then, the action of the G-CAT model is the WZNW action for the field $B_0\in
\exp\left(\cgh_{0}^{\ast}\right)$, plus a potential involving the field
$B_0$ coupled to $\eta$, plus, finally, a kinetic term for the two fields
$\nu$ and $\eta$. Consequently, when $\eta=0$ (or constant, in general) the
action \rf{brokenwznwplus} is precisely the action of the generalized
non-abelian affine Toda model corresponding to the field $B_0$.

Eq.\rf{brokenwznwplus} also shows that one should not expect to obtain always
different models for different values of $l$. First, notice that the explicit
dependence on $l$ in $e^{l\eta}$ can be eliminated by $\eta\rightarrow \eta/l$
and $\nu\rightarrow \nu\>l$. Second, it follows from
eqs.\rf{choicegrad}-\rf{liegrad} that
\be
\cgh= \IR\>C\> \oplus \IR\>D\> \bigoplus_{j\in E}\> \cgh_j,
\ee
where $E= I+ \IZ\> N_{\bf s}$, $I$ is a set of integers $\geq0$ and
$<N_{\bf s}$, and $\cgh_{j+ m\>N_{\bf s}}$ is isomorphic to $\cgh_j$ for all
$m\in \IZ$. Through this isomorphism, for each $u\in \cgh_j$ and $v\in
\cgh_{-j}$ such that
$Tr(u\>v)\not=0$ there exists $\tilde u\in \cgh_{j+m\>N_{\bf s}}$ and $\tilde
v\in \cgh_{-j-m\>N_{\bf s}}$, for all $m\in \IZ\geq0$, such that $Tr(u\>v)
= Tr(\tilde u\>\tilde v)$. All this shows that the only {\it a priori}
different
models correspond to the different choices of $\Lambda_{\pm l}$ with $0<l\leq
N_{\bf s}$.

The symmetry group of the G-CAT model is the subgroup of the symmetry group of
the WZNW model \rf{symwznw} which leaves the potential term invariant.
Specifically, \rf{brokenwznw} is invariant under the transformations
\br
B(x_{+},x_{-}) \ra h_L(x_{-}) B(x_{+},x_{-}) \qquad , \qquad
B(x_{+},x_{-}) \ra B(x_{+},x_{-}) h_R(x_{+})
\lab{brokengauge}
\er
where $h_{L/R}$ are elements of certain subgroups $\ch_{L/R}\subset
\exp\left(\cgh_0\right)$ satisfying
\be
 h_L^{-1} \Lambda_{-l}  h_L = \Lambda_{-l} \qquad , \qquad
h_R \Lambda_{l}  h_R^{-1} = \Lambda_{l}
\lab{brokeninv}
\ee
Depending upon the choice of $\Lambda_{\pm l}$, these subgroups may vary
considerably, and, in fact, the left and right subgroups might not be
isomorphic. The Noether currents associated to such symmetries are
\be
 J(T_L) = -\Tr \( T_L \pa_{-}B B^{-1} \)\, ; \qquad
J(T_R) = \Tr \( T_R B^{-1}\pa_{+}B \)
\ee
where
\be
T_{L/R} \in {\rm Ker\/}\left( {\rm ad\;} \Lambda_{\mp l}\right) \cap \cgh_0,
\ee
{\it i.e.\/},
\be
\lb T_L \, , \, \Lambda_{-l} \rb = 0 \,, \quad
\lb T_R \, , \, \Lambda_l \rb = 0\,, \quad {\rm and}\quad
\exp\left( T_{L/R}\right) \in \ch_{L/R} .
\ee
Indeed, from \rf{natoda1} and \rf{natoda2} we have
\be
\pa_{+} J(T_L) = 0 \, ; \qquad  \pa_{-} J(T_R) = 0
\ee

The question of the vacua of these models is quite delicate because the sign
of the kinetic energy depends on the properties of the trace form $\Tr$.
Actually, we will be interested in the solitons of the
massive G-AT models~\rf{gat}, which correspond to the spontaneous breakdown of
the conformal symmetry by the choice $\eta=\eta_0={\rm constant}$ (see
\rf{higgs}). Then, the only field will be $B_0$, and eq.\rf{brokenwznwplus}
shows that the sign of the kinetic term depends only on the
restriction of the trace form to $\cgh_{0}^{\ast}$ or, equivalently, on the
properties of the subgroup $\exp\left(\cgh_{0}^{\ast}\right)$ regarding its
compactness. Even though affine Toda models based on non-compact groups have
been considered in relation to two-dimensional black holes~\ct{gervais}, we
will be mainly interested in the case when the group is compact and the kinetic
energy of $B_0$ is positive. This can be achieved by choosing the compact real
form of the Lie algebra $\cgh_{0}^{\ast}$, which is always possible because of
the following reason. Let us recall that
$\cgh_{0}^{\ast}$ is isomorphic to the affine subalgebra of $\cgh$ generated by
$\{H_{1}^0,\ldots,  H_{r}^0\}$\footnotemark{\footnotetext{Actually, the
generators are
$\{\widetilde{H}_{1}^0,\ldots,
\widetilde{H}_{r}^0\}$, but, because of equation \rf{bilbasis2}, the $H_a$'s
are equivalent to the $\widetilde{H}_a$'s as generators of the subalgebra
$\cgh_{0}^{\ast}$.}}, and $\{E_{\alpha}^n\}$ with $n\>N_{\bf s}+\alpha\cdot
H_{\bf s}=0$ ---in appendix~\ref{ap:gcero}, we show that the only
possible values of $n$ are either $0$, if $s_0\not=0$, or $0,\pm1$, if
$s_0=0$---. Then, if
$E_{\alpha}^n\in \cgh_{0}^{\ast}$ for some $\alpha$ and
$n$, $E_{-\alpha}^{-n}\in \cgh_{0}^{\ast}$ too, and we can always choose
the real form of $\cgh_{0}^{\ast}$ such that its generators are $H_1^0, \ldots,
H_r^0$, $(E_{\a}^n + E_{-\a}^{-n})$ and $i(E_{\a}^n - E_{-\a}^{-n})$, where
$n\>N_{\bf s}+\alpha\cdot H_{\bf s}=0$. With this choice,
eqs.\rf{trace1}-\rf{trace3} can be used to show that the trace form of $\cgh$
restricted to
$\cgh_{0}^{\ast}$ is positive definite; this is equivalent to say that the
subgroup $\exp\left( \cgh_{0}^{\ast} \right)$ is compact. In the previous
discussion, we have ignored the field $\nu$ because it does not
contribute to the kinetic energy of the G-AT model, as can be checked in
\rf{brokenwznwplus}. Thus, for $\eta = \eta_0= {\rm constant}$, the action
\rf{brokenwznwplus} has a positive definite kinetic energy for the
particular choice of the real form of
$\cgh_{0}^{\ast}$.

In addition, with this choice, $B_0$ has to be unitary, which ensures the
reality of the first two terms of \rf{brokenwznwplus} that specify a particular
WZNW model. In contrast, the reality of the potential depends on the
hermiticity properties of $\Lambda_{\pm l}$, which strongly constrains the
possible choices of $\Lambda_{\pm l}$~\ct{tlp}. However, as we discuss in
sections~\ref{sec:classmass} and \ref{sec:solmass}, the masses of the
fundamental particles and solitons of these models depend only upon the
eigenvalues of certain elements $E_{\pm l}\in \cgh_{\pm l}$ defined in
\rf{epml}, and those masses are real even for a wide class of models
whose potential is complex.

In order to have static solutions for the field $B_0$, the right
hand side of~\rf{eqb} should vanish or have components in the direction of the
central term only, {\it i.e.\/}, we will be interested in those cases where
there exists a constant group element $b_0\in \cgh_{0}$ such that
\be
E_{l} \equiv b_0 \Lambda_{l} b_0^{-1} \quad {\rm and}\quad
E_{-l} \equiv \Lambda_{-l}
\lab{epml}
\ee
satisfy
\be
\lb  E_{l} \, , \, E_{-l} \rb = \b\> C,
\lab{vacua}
\ee
where
\be
\b \> =\> {1\over N_{\bf s}}\> \Tr\left( \qs[E_l, E_{-l}]\right)\>
=\> {l\over N_{\bf s}}\> \Tr\left(E_l\>E_{-l}\right).
\ee
Notice that eq.\rf{vacua} would be satisfied if $E_{\pm l}$ are elements
of some Heisenberg subalgebra of $\cgh$; this shows that it is possible to
associate G-CAT models that admit static solutions to
the elements of the different Heisenberg subalgebras of $\cgh$ whose
classification is available~\ct{kacpet}. Then, for $\eta=0$, the
corresponding solution of~\rf{eqb} is
\be
B_{0}^{\rm vac} = b_0 \quad {\rm and}\quad \nu_0 \,=\, -\>\b\> x_+\> x_-
\lab{b0vac}.
\ee
Obviously, if $\b \neq 0$, the vacuum solution for the  field in the direction
of the central term $C$ is not actually static; in fact, this is
a generalization of what occurs in the abelian CAT models~\ct{acfgz}.

Let us introduce a group element $b^{\ast}\in \cgh_{0}^{\ast}$ such that (see
eq.~\rf{parb})
\be
B = B_0 e^{\nu \>C\> +\>\eta\>\qs}  = b\> b_0 \, e^{\nu_0\> C} e^{\eta \>
\qs}=   b^{\ast}\>e^{\nu\> C}\> b_0 \, e^{\nu_0\> C} e^{\eta \> \qs};
\lab{bbar}
\ee
so, the vacuum solution is $b^{\ast}=1$, $\nu=0$, and $\eta =0$. Therefore, the
equations of motion
\rf{natoda1} read
\be
\pa_{+} \( \pa_{-} b^{\ast} \, b^{\ast-1} \) +(\pa_{+}\pa_{-}\nu - \b)\> C +
\pa_{+}\pa_{-}
\eta
\> \qs = e^{l\, \eta}\>\lb E_{-l}\, ,
\, b^{\ast} E_{l} b^{\ast-1} \rb;
\lab{gcatnice}
\ee
obviously, according to \rf{eta}, $\eta$ is a free field.
In terms of the field $b^{\ast}$, the transformations
\rf{brokengauge} read
\be
b^{\ast}\>e^{\nu\> C} \ra {\tilde h}_L (x_{-})  b^{\ast}\>e^{\nu\> C} \, ;
\qquad   b^{\ast}\>e^{\nu\> C} \ra   b^{\ast}\>e^{\nu\> C}
\, {\tilde h}_R (x_{+})
\lab{brokeninv2}
\ee
where
\be
{\tilde h}_R (x_{+})\equiv b_0 h_R(x_{+}) b_0^{-1}, \qquad
{\tilde h}_L (x_{+})\equiv h_L(x_{+}),
\lab{newhr}
\ee
and so from \rf{brokeninv} and \rf{epml}
\be
{\tilde h}_L(x_{-})^{-1}\, E_{-l}\, {\tilde h}_L(x_{-}) =  E_{-l} \quad
{\rm and} \quad
{\tilde h}_R (x_{+}) \, E_l \, {\tilde h}_R (x_{+})^{-1} = E_l.
\lab{invepml}
\ee

In addition to the continuous symmetries expressed
by eqs.\rf{brokeninv2}-\rf{invepml}, the theory may possess some discrete
symmetries. In general, these are generated by group elements which are
exponentiations of generators that do not really commute with $E_{\pm l}$, but
do leave them invariant when the parameters in the exponentiation assume
special values. An interesting class of such symmetries   is related to the
co-weight lattice of
$\cg$. For the integral gradations
\rf{niceq}, the Cartan subalgebra of $\cgh_0$ is the same as that of $\cgh$.
Then, let $\mu^v$ be a co-weight of $\cg$, {\it i.e.\/}, $\mu^v = \sum_{a=1}^r
m_a 2 \l_a /\a_a^2$, with $m_a$ being integers,  and $\l_a$ and $\a_a$ being
the
fundamental weights and simple roots of $\cg$, respectively. Any element of the
Kac-Moody algebra $\cgh$ has an integer eigenvalue with respect to the elements
of the Cartan subalgebra of the form
$\mu^v
\cdot H^0 + n\,D$; therefore
\be
e^{2 \pi i (\mu^v \cdot H^0+n\, D)} E_{\pm l} e^{-2 \pi i (\mu^v \cdot
H^0+n\, D)} =
E_{\pm l} \>, \qquad \mbox{\rm for any co-weight $\mu^v$ and integer $n$}
\lab{degenerate}
\ee
Obviously, this is a complex transformation; however, since we will be dealing
with the compact real form of $\exp\left(\cgh_{0}^{\ast}\right)$, it implies a
real transformation on the components of the field $b$, when we parameterize
it as in \rf{bbar2}. Such class of transformations constitute a generalization
of what occurs in the abelian affine Toda models, in the pure imaginary
coupling constant regime~\ct{hollo,acfgz,otu}. Those transformations are
responsible for the infinitely many degenerate vacua and for the existence of
topological solitons. In addition to \rf{degenerate}, some  theories may
possess some additional discrete symmetries depending upon the particular form
of $E_{\pm l}$. The  question of the topological charges associated to
such discrete symmetries will be discussed elsewhere.

\sect{The classical masses of the fundamental particles}
\label{sec:classmass}

\ \indent
The G-CAT models, described by \rf{brokenwznw}, are conformally invariant and,
therefore, their fundamental particles are massless. However, the G-AT models
introduced in \rf{gat} are the gauge fixed version of the G-CAT models,
which are not conformally invariant, and which have massive fundamental
particles; in this section, we will calculate their classical masses. In
\rf{bbar}, we will parameterize the group  element
$b^{\ast}\in \cgh_{0}^{\ast}$ as
\be
b^{\ast} \equiv e^{i\, T}
\lab{bbar2}
\ee
where
\be
T\equiv \vp^i \, T_i\>, \qquad i\>=\> 1,2,\ldots, {\rm dim}\,
\cgh_{0}^{\ast}\>,
\ee
where ${\rm dim}\, \cgh_{0}^{\ast}={\rm dim}\, \cgh_0 -2$, and $T_i$ denote all
the generators of $\cgh_{0}^{\ast}$ (see \rf{paramalg}); $\{\vp^i\}$ and $\nu$
are fields of the the G-AT model. Then, taking $\eta=0$, the linear part of the
eq.\rf{gcatnice} gives the  Klein-Gordon equations for the G-AT fields
\br
\pa_{+}\pa_{-} \nu &=& 0
\lab{kg2}\\
\pa_{+}\pa_{-}\, \vp^i \, T_i + \vp^i \,\lb E_{-l}\, , \,\lb E_l \, , \,
T_i\rb\rb &=&  0
\lab{kg1}
\er
So, the field $\nu$ is massless, and, writing \rf{kg1} as (our conventions are
$x_{\pm}=x\pm t$, and $\pa^2 = -4 \pa_{+}\pa_{-} = \pa_{t}^{2}-\pa_{x}^{2}$)
\be
\(\pa^{2}\, \vp^j + \vp^i \, {M^2}_{i}^{j} \)\, T_j = 0\>,
\lab{kge}
\ee
we get that the mass square matrix for the $\vp^i$ fields is given by
\be
{M^2}_{i}^{j} \, T_j \equiv - 4 \lb E_{-l}\, , \,\lb E_l \, , \,
T_i\rb\rb\> = \>- 4 \lb E_{l}\, , \,\lb E_{-l} \, , \, T_i\rb\rb,
\lab{massmatrix2}
\ee
where we have used eq.~\rf{vacua}; therefore
\br
M_{ij}^2 \equiv {M^2}_{i}^{k}\, \Tr \( T_k \, T_j\)&=&
-2 \(\Tr \(\lb E_{-l}\, , \,\lb E_l \, , \, T_i\rb\rb T_j\)
+ \Tr \(\lb E_{-l}\, , \,\lb E_l \, , \, T_j\rb\rb T_i\)\)
\nonu\\
&=& -2 \( \Tr\(\lb  E_{-l} \, , \,  T_j \rb \lb T_i\, ,\, E_{l} \rb\)
 + \Tr\( \lb E_{-l} \, , \, T_i \rb\lb T_j\, ,\, E_{l} \rb\) \)
\lab{massmatrix}
\er
If ${\tilde h}_L(x_-) $ and ${\tilde h}_R (x_{+})$, defined in \rf{invepml},
have a set of common generators, one observes from the second line in
\rf{massmatrix} that the mass matrix is block diagonal with the block
corresponding to this set  being zero; so, the particles associated to that
set of common generators are massless.

Notice that the diagonalization of the mass matrix \rf{massmatrix2} corresponds
to the diagonalization of the action of the operator $\lb E_{-l} \, , \lb E_{l}
\, , \, \cdot \rb\rb$ on the subalgebra $\cgh_{0}^{\ast}$,
\be
\lb E_{-l} \, , \lb E_{l}\, , \, T \rb\rb =
\lb E_{l} \, , \lb E_{-l}\, , \, T \rb\rb =\l T \, \, ; \qquad T \in
\cgh_{0}^{\ast}
\lab{masseigen}
\ee
Due to \rf{vacua} and to the invariance of the trace form, the subspaces
corresponding to different eigenvalues are orthogonal. Therefore, the masses
of the fundamental particles are
\be
m_{\l}^2 = -4 \,  \l.
\lab{partmass}
\ee
For the G-CAT models, this constitutes the generalization of the arguments used
in the abelian affine Toda models~\ct{freeman,braden}. However,
to establish the physical significance of \rf{partmass} in each G-CAT model,
one has to ensure that the eigenvalues $\l$ are real and non-positive. Below,
we
discuss how to calculate the eigenvalues $\l$ in some examples.

When going from the G-CAT to the G-AT model, notice that if we had set the
field
$\eta$ to a non-vanishing constant value, namely $\eta=\eta_0$, the commutator
term in \rf{gcatnice} would get multiplied by a factor
\be
v_{\eta} \equiv e^{l \eta_0}
\lab{higgs}
\ee
and all the masses would be multiplied by this
factor too, $m_{\l}^2 = -4 \, v_{\eta}\, \l$. Therefore, the primary field
$\phi
\equiv e^{l \,\eta}$~\rf{primary} actually works like a Higgs field with the
mass matrix being proportional to its vacuum expectation value $v_{\eta}$.
Consequently, the masses of the G-AT model are generated by the spontaneous
breakdown of the conformal invariance of the G-CAT models by the choice of a
particular vacuum configuration $\eta=\eta_0 = {\rm constant}$.  However, we
will not carry the constant $v_{\eta}$ explicitly because it can be absorbed in
the definition of $E_{\pm l}$; so, the  mass scale of the G-AT models will be
set by the normalization of $E_{\pm l}$.

\sect{The Dressing Transformations}
\label{sec:dressing}

\ \indent
Consider a generic non-linear system admitting a formulation in terms of a
zero-curvature condition
\be
\lb \pa_{\mu} + A_{\mu} \, , \, \pa_{\nu} + A_{\nu}\rb = 0
\lab{zcdress}
\ee
with the gauge potentials $A_{\mu}$ lying on a Lie algebra $\cgh$ and being
functionals of the fields of the model. The dressing transformations
\ct{dress1,dress2,dress3,dress4} are non-local gauge transformations acting on
$A_{\mu}$ and leaving its form invariant. Each one of these gauge
transformations is performed in two different ways in terms of two different
group elements $\tpp$ and $\tmm$ of $G$ (the group whose Lie algebra is $\cgh$)
\be
 A_{\mu} \ra  A_{\mu}^g \equiv \Theta_{\pm}  A_{\mu} \Theta_{\pm}^{-1} -
\pa_{\mu} \Theta_{\pm} \Theta_{\pm}^{-1}
\lab{dressing}
\ee
Since $A_{\mu}$ satisfies \rf{zcdress} it has to be of pure gauge form
\be
A_{\mu} = - \pa_{\mu} T T^{-1}
\lab{puregauge}
\ee
and, consequently, $A_{\mu}^g = - \pa_{\mu} (\Theta_{\pm}T)
(\Theta_{\pm}T)^{-1}$. Therefore, there exists a constant group element $g$
such that $\tpp T = \tmm T g$ and, so,
\be
\tpp T = \tmm T g\quad  {\rm or}\quad {\rm equivalently}\quad \tmm^{-1}
\tpp = T g T^{-1}
\lab{reltptm}
\ee

The requirement that the dressing transformation preserves the form of the
potential $A_{\mu}$ implies that $\Theta_{\pm}$ must  belong to two different
subgroups of $G$ determined by the particular form of $A_{\mu}$ in $\cgh$. In
this sense, eq.~\rf{reltptm} corresponds to the factorization of $T g T^{-1}$
into those subgroups. Then, given a solution of the model defined by a
group element $T$ and a constant group element $g$, the dressing transformation
can be used to construct another  solution defined by $ T^g =\tpp
T$ (or equivalently  $T^g= \tmm T$).

We will now use the dressing transformations for the G-CAT models to construct
their solutions lying in the orbit of the vacuum. The gauge potentials for the
G-CAT models are given by \rf{gp}. Using  \rf{epml} and \rf{bbar}
one gets
\br
A_{+} &=& - e^{l \, \eta}\, b E_l b^{-1} \nonu\\
A_{-} &=& - \pa_{-} b b^{-1} - \pa_{-} \eta \, \qs  + E_{-l} + \b x_{+} \, C.
\lab{gp2}
\er
Then, for the vacuum solution, namely $b=b^\ast =1$ and $\eta = 0$, one has
\br
A_{+}^{(0)} &=& -  E_l \nonu\\
A_{-}^{(0)} &=&   E_{-l} + \b x_{+} \, C
\lab{vacgp}
\er
which can be written as in \rf{puregauge} with the group element
\be
T_0 = e^{x_{+} \, E_l} \,e^{-x_{-} \, E_{-l}}
\lab{t0}
\ee
We now perform a gauge transformation of the form~\rf{dressing} that maps
\rf{vacgp} into \rf{gp2} for $\eta = 0$. Due to the structure of $A_{\pm}$, we
choose the elements $\Theta_{\pm}$ in \rf{dressing} as
\be
\tpp = \tp0 \,\tpg \, ; \qquad \qquad \tmm = \tm0 \,\tms
\lab{tpm}
\ee
with $\Theta_{\pm}^{(0)}$, $\tpg$ and $\tms$ belonging to the
subgroups of $G$ obtained by exponentiating the subalgebras $\cgh_0$,
$\cgh_{+}$ and $\cgh_{-}$, defined in \rf{grade3}, respectively; consequently,
we write
\br
\tpg = \exp \( \sum_{s>0} t^{(s)}\) \, ; \qquad
\tms = \exp \( \sum_{s>0} t^{(-s)}\) \, ;
\qquad \qquad t^{(\pm s)} \in \cgh_{\pm s}
\lab{tpms}
\er
Therefore $\Theta_{\pm}$ have to satisfy
\br
 b E_l b^{-1} &=& \Theta_{\pm} E_l \Theta_{\pm}^{-1} + \pa_{+} \Theta_{\pm}
\Theta_{\pm}^{-1}
\lab{cond1}\\
E_{-l} - \pa_{-}b b^{-1}&=& \Theta_{\pm}E_{-l} \Theta_{\pm}^{-1} - \pa_{-}
\Theta_{\pm} \Theta_{\pm}^{-1}
\lab{cond2}
\er
In order to solve these equations one has to split them into the
eigensubspaces of the gradation \rf{grade}. We now show  how to construct
$\tpp$. The calculations for $\tmm$ are very similar. Due to the structure
of $\tpp$ the right hand side of \rf{cond1} and \rf{cond2} have grades greater
or equal to $0$ and $-l$ respectively. So, using \rf{tpm} one gets that the
component of grade zero in \rf{cond1} implies $\pa_{+} \tp0 (\tp0 )^{-1} =0$
and the component of grade $-l$ of \rf{cond2} implies $E_{-l}= \tp0 E_{-l}(\tp0
)^{-1}$. Therefore, $\tp0$ has to be an element of the subgroup ${\cal H}_L$
introduced in \rf{brokengauge}. So, we write
\be
\tp0 \equiv h_L^{-1} (x_{-})
\lab{tp0}
\ee

Considering the component of \rf{cond1} with grades $0<s<l$, one gets
\be
\(\pa_{+} \tpg (\tpg )^{-1}\)_{0<s<l} = 0\, , \nonu
\ee
and from the component of \rf{cond2} of grades $-l<s<0$ it follows that
\be
\(\tpg E_{-l}(\tpg)^{-1}\)_{-l<s<0}=0\, . \nonu
\ee
Therefore, taking into account \rf{tpms}, one concludes that
\be
t^{(s)} \equiv t^{(s)}(x_{-}) \in \mbox{\rm centralizer of $E_{-l}$, for
$0<s<l$}
\lab{tpchiral}
\ee
Now, using \rf{tp0} and \rf{tpchiral}, the component of grade $l$ of
eq.\rf{cond1} leads to
\be
\pa_{+} t^{(l)} = - E_l + h_L (x_{-}) b E_l b^{-1}  h_L^{-1} (x_{-})
\lab{tpl1}
\ee
As for the component of grade $0$ in \rf{cond2} one gets
\be
\pa_{-} b b^{-1} = - h_L^{-1} (x_{-})\pa_{-} h_L (x_{-})  -
h_L^{-1} (x_{-}) \lb t^{(l)}\, , \, E_{-l} \rb h_L (x_{-})
\lab{tpl2}
\ee
Finally, for the components of \rf{cond1} and \rf{cond2} of grades greater than
$l$ and $0$ respectively, the result is
\br
\(\pa_{+} \tpg (\tpg )^{-1} + \tpg \, E_l \, (\tpg )^{-1} \)_{>l} &=& 0
\lab{tpgl}\\
\(\pa_{-} \tpg (\tpg)^{-1} - \tpg \, E_{-l} \, (\tpg )^{-1} \)_{>0} &=& 0
\lab{tpg0}
\er

Performing the same calculation for $\tmm$, one finds
\br
\tm0 &=& b \, {\tilde h}_R (x_{+})
\lab{tm0}\\
t^{(-s)} &=& t^{(-s)}(x_{+}) \in \mbox{\rm centralizer of $E_{l}$, for
$0<s<l$}
\lab{tmchiral}\\
\pa_{-} t^{(-l)} &=&  E_{-l} - {\tilde h}_R (x_{+})^{-1}  b^{-1} E_{-l} b
{\tilde h}_R (x_{+})
\lab{tml1}\\
b^{-1} \pa_{+} b &=& - \pa_{+} {\tilde h}_R (x_{+}){\tilde h}_R (x_{+})^{-1} -
{\tilde h}_R (x_{+}) \lb t^{(-l)} \, , \, E_l \rb {\tilde h}_R (x_{+})^{-1}
\lab{tml2}
\er
where ${\tilde h}_R (x_{+})$ was defined in \rf{newhr}, and also
\br
\(\pa_{+} \tms (\tms )^{-1} + \tms \, E_l \, (\tms )^{-1} \)_{<0} &=& 0
\lab{tms0}\\
\(\pa_{-} \tms (\tms )^{-1} - \tms \, E_{-l} \, (\tms )^{-1} \)_{<-l} &=& 0
\lab{tmsl}
\er

Taking all this into account, from \rf{reltptm}, one obtains
\br
(\tms )^{-1} \( h_L(x_{-}) \, b \, {\tilde h}_R (x_{+})  \)^{-1} \tpg &=&
T_0 \, g T_0^{-1} \nonu\\
&=& e^{x_{+} \, E_l} \,e^{-x_{-} \, E_{-l}} \, g \, e^{x_{-} \, E_{-l}} \,
e^{-x_{+} \, E_l}
\lab{important}
\er
This relation has a very powerful meaning and it will be important in
understanding the soliton solutions and $\tau$-functions. Once $\tpg$ and
$\tms$ are  determined from the relations given above, it is possible to
reconstruct the corresponding solution for the $b$ fields on the orbit of the
vacuum for each constant group element $g$. In~\rf{important}, notice that he
gauge symmetries \rf{brokengauge} of the G-CAT models have been made explicit,
and one can easily choose a unique solution in each one of the orbits of those
gauge groups.

Since we will use it in section~\ref{sec:examples}, we point out here that
the choice
\be
t^{(n)}=t^{(-n)}=0\, ; \qquad \mbox{\rm for $n$ not a multiple of $l$}
\lab{easier}
\ee
for the quantities introduced in~\rf{tpms}
is a solution for the dressing problem that corresponds to choosing all the
chiral quantities in eqs.\rf{tpchiral} and~\rf{tmchiral} to be zero. Indeed
\rf{tpchiral} and
\rf{tmchiral} are trivially satisfied and \rf{tpgl}, \rf{tpg0}, \rf{tms0} and
\rf{tmsl} will now have components only in the subspaces with grade which are
multiples of
$l$, and  so, they can be consistently solved. However, unless some
extraordinary cancellation happens, the type of solutions \rf{easier} imply
that the constant group element $g$ in \rf{important} can only be an
exponentiation of generators with grades which are multiples of $l$.

\subsection{Connection with the Leznov-Saveliev solution}

If $\mid \mu \rangle$ and $\mid \mu^{\pr} \rangle$ are states of a given
representation of $\cgh$ satisfying \rf{hws} and \rf{lws}, and after fixing the
gauge as ${\tilde h}_R=h_L=1$, one obtains from
\rf{important} that
\br
\langle \mu^{\pr} \mid b^{-1} \mid \mu \rangle &=& \langle \mu^{\pr} \mid
b_0 B^{-1} e^{\nu_0\> C} \mid \mu \rangle \nonu\\
&=& e^{-\b x_{+}x_{-} \, C(\mu )}
\langle \mu^{\pr} \mid e^{x_{+} \, E_l}  \, g \, e^{x_{-}
\, E_{-l}} \mid \mu \rangle
\lab{vacorbit}
\er
where we have used \rf{vacua}, and $C(\mu )$ is the value of the central term
in
the representation under consideration.

Now, if one chooses the parameters of the Leznov-Saveliev solution
\rf{solution} as
\be
B_{+} = b_0^{-1} \qquad \qquad B_{-} = 1
\lab{choice1}
\ee
and
\be
\chi^{-}_{s}=\chi^{+}_{-s}=0 \, ; \qquad \qquad \mbox{\rm for $0<s<l$}
\lab{choice2}
\ee
then one gets from \rf{eqform} and \rf{eqforn} that
\be
M_{-}(x_{-}) = e^{-x_{-} \, E_{-l}}\, M_{-}(0)
\, ; \qquad \qquad
N_{+}(x_{+}) = e^{x_{+} \, E_{l}} \, N_{+}(0)
\ee
Substituting that into \rf{solution}, and using \rf{parb} and \rf{bbar}, we get
the same as \rf{vacorbit} with $g\equiv N_{+}(0) M_{-}(0)^{-1}$. The factor
$e^{\eta \, \qs}$ drops out if the state $\mid \mu \rangle$ has zero
$\qs$-eigenvalue, and a factor $b_0^{-1}$ on both sides of the relation can be
eliminated before taking the expectation value on the representation states.

Therefore the quite simple choice of parameters \rf{choice1} and \rf{choice2}
leads, in the Leznov-Saveliev method, to all the solutions in the orbit, under
the dressing transformations, of the vacuum solution. In addition, such choice
picks up only one solution in each orbit of the gauge transformations
\rf{brokengauge}.  Such observation sheds some light on the so called {\em
solitonic specialisation procedure} used in \ct{otu} to construct the
soliton solutions of the abelian Affine Toda models, and suggested in~\ct{osu}
for the same construction, in the case of a more general class of integrable
models.

\sect{The $\tau$-functions}
\label{sec:tau}

\ \indent
In relation \rf{important} the solution for the field $b$ is found in terms
of the quantities $t^{(s)}$. In practice, solving such relations might turn in
a quite cumbersome task. However, the quantities $t^{(s)}$ have a very
important meaning. They are the embryo of the so called $\tau$-functions.

The $\tau$-functions have a very important role in soliton theory. They are
used in the Hirota method to construct exact solutions of soliton equations
\ct{hiro}. In addition, formalisms have been developed to construct
soliton equations starting from a given definition of the $\tau$-function
\ct{miwa,kw}. The connection between dressing transformations and
$\tau$-functions has been explored in the context of the KdV
equation~\ct{wilson} and of the generalized integrable hierarchies of the KdV
type~\ct{tault}. However, in those cases the $\tau$-functions were
defined on some special vertex representation of simply laced Kac-Moody
algebras. Inspired on the dressing transformations, here, we present a
general definition of the $\tau$-function that is not restricted to any
particular representation of $\cgh$, and which generalizes the results
of~\ct{tault}.

Consider an integral gradation of an affine Kac-Moody
algebra $\cgh$ labelled by a vector $\bf s$~\rf{niceq}, and let $\mid \l_{\bf
s} \rangle$ be the highest weight state  of an integrable representation of
$\cgh$
\ct{kac1}, satisfying the relations
\rf{int0}-\rf{int4}. We define the  $\tau$-function $\tau_{\bf s}$
as the result of the action of the left hand side of \rf{important} (or
equivalently, its right hand side) on  $\mid \l_{\bf s} \rangle$, but with the
gauge symmetry \rf{brokengauge} fixed as $h_L(x_{-})={\tilde h}_R(x_{+})=1$,
{\it i.e.\/},
\br
\tau_{\bf s}\, (x_{+}\, , \, x_{-}) &\equiv&
(\tms )^{-1}  \, b^{-1}\, \mid \l_{\bf s} \rangle \nonu\\
&=& e^{x_{+} \, E_l}\, e^{-x_{-} \, E_{-l}}\, g \, e^{x_{-} \, E_{-l}}
\mid \l_{\bf s} \rangle
\lab{tau}
\er
As far as the relation among the $b$ fields and $\tau$-functions is
concerned,  the definition above is made on shell. However once such relation
is established it can be extended off shell.

The definition \rf{tau} implies that
\br
\pa_{+} \tau_{\bf s} &=& \lb E_l \, , \, G \rb  \, \mid \l_{\bf s} \rangle
= E_l \, \tau_{\bf s}
\lab{eqtau1}\\
\pa_{-} \tau_{\bf s} &=& - \lb E_{-l}\, , \, G \rb \, \mid \l_{\bf s} \rangle
\lab{eqtau2}
\er
where we have used \rf{vacua} and denoted
\be
G \equiv e^{x_{+} \, E_l}\, e^{-x_{-} \, E_{-l}}\, g \, e^{x_{-} \, E_{-l}}
\, e^{-x_{+} \, E_l} = T_0 \, g \, T_0^{-1}
\ee
Therefore,
\be
\pa_{+}\pa_{-} \tau_{\bf s} = - \lb E_l \, , \lb E_{-l} \, , G \rb\rb \, \mid
\l_{\bf s} \rangle,
\lab{eqtau3}
\ee

Since $\tau_{\bf s}$ is a state of the representation, we can split it into
eigenstates of the grading operator $Q_{\bf s}$ given by \rf{gradop}
\be
\tau_{\bf s} = \tau_{\bf s}^{(0)} + \tau_{\bf s}^{(-1)} + \tau_{\bf s}^{(-2)} +
\ldots
\lab{modeexp}
\ee
with
\be
Q_{\bf s} \, \tau_{\bf s}^{(n)}= (\eta_{\bf s} + n) \, \tau_{\bf s}^{(n)},
\ee
where $\eta_{\bf s}$ has been defined in~\rf{int3}.

Notice that $\mid \l_{\bf s} \rangle$ is an eigenstate of the subalgebra
$\cgh_0$ and, since $\tms$ is an exponentiation of the negative grade
operators, we get that
\be
\tau_{\bf s}^{(0)} = b^{-1} \, \mid \l_{\bf s} \rangle \sim \, \mid
\l_{\bf s} \rangle
\lab{tau0}
\ee
Thus, denoting
\be
{\hat \tau}_{\bf s}^{(0)} \equiv \langle \l_{\bf s}\mid \tau_{\bf s}^{(0)} =
\langle \l_{\bf s}\mid \tau_{\bf s} = \langle \l_{\bf s}\mid \, b^{-1} \, \mid
\l_{\bf s} \rangle
\lab{tauhat0}
\ee
we get
\be
{\tau_{\bf s} \o {\hat \tau}_{\bf s}^{(0)}} = (\tms )^{-1}\, \mid \l_{\bf
s} \rangle.
\lab{nuestra}
\ee
This equation is the straightforward generalization of the eq.(5.1)
of~\ct{tault}, which, in that reference, establishes the relation between the
zero curvature formalism and the $\tau$-functions for a class of generalized
integrable hierarchies of partial differential equations; our results open the
possibility of generalizing~\ct{tault} for a wider class of integrable
equations. As a consequence of
\rf{eqtau1}, notice that the c-number function ${\hat
\tau}_{\bf s}^{(0)}$ satisfies
\be
\pa_{+} \ln {\hat \tau}_{\bf s}^{(0)} =   \langle \l_{\bf s}\mid \, E_l \,
(\tms )^{-1} \, \mid \l_{\bf s} \rangle
\ee

Solutions that travel with constant velocity without dispersion, namely the
one-soliton solutions, can be easily constructed as follows. Let $F$ be an
eigenvector under the adjoint action of $E_{\pm l}$
\be
\lb E_{\pm l} \, , \, F \rb = \omega_{\pm l} F
\lab{epmleigen}
\ee
We then choose the constant group element $g$ in \rf{tau} as
\be
g_{\rm sol.} \equiv e^F
\lab{gsol}
\ee
Therefore
\br
\tau_{\bf s}^{\rm 1-sol.}\, (x_{+}\, , \, x_{-}) &\equiv&
e^{x_{+} \, E_l}\, e^{-x_{-} \, E_{-l}}\, e^F \, e^{x_{-} \, E_{-l}}
\mid \l_{\bf s} \rangle\nonu\\
&=& \exp \( e^{\gamma ( x - vt)} F\) \, \mid \l_{\bf s} \rangle
\lab{onesol}
\er
where $x_{\pm}=x\pm t$, and
\be
\gamma = \omega_{l} - \omega_{-l} \, ; \qquad \qquad
 v =- { {\omega_{l} + \omega_{-l}}\o{\omega_{l} - \omega_{-l}}}
\lab{solpar}
\ee
Nevertheless, if $v$ is going to be the velocity of the soliton, observe that
we
must have
\be
\omega_{l}\,\omega_{-l} <0
\ee
because if $\omega_{\pm l}$ have the same sign, then $\mid v \mid >1$.

So, for each eigenvector of $E_{\pm l}$ we have a one-soliton solution. The
multi-soliton solutions are constructed by choosing $g$ as a product of
one-soliton $g$'s, {\it i.e.\/}, $g=e^F \, e^{F^{\pr}}\, e^{F^{\pr\pr}}
\ldots$. In fact, this is a generalisation of the ideas used in \ct{otu}.
However, notice that if we have $F$ and $F^{\pr}$, with corresponding
eigenvalues $\omega_{\pm l}$ and $\omega_{\pm l}^{\pr}$, satisfying \rf{solpar}
with different $\gamma$'s (let's say $\gamma$ and $\gamma^{\pr}$) but the same
velocity $v$, then we have
\br
\tau_{\bf s}^{\rm 2-sol.}\, (x_{+}\, , \, x_{-}) &\equiv&
e^{x_{+} \, E_l}\, e^{-x_{-} \, E_{-l}}\, e^F\, e^{F^{\pr}} \, e^{x_{-} \,
E_{-l}}
\mid \l_{\bf s} \rangle\nonu\\
&=& \exp \( e^{\gamma ( x - vt)} F\) \, \exp \( e^{\gamma^{\pr} ( x - vt)}
F^{\pr}\)\,\mid\l_{\bf s} \rangle
\lab{sta2sol}
\er
So, we have a two-soliton solution which can be put at rest in some Lorentz
frame. Such solution possesses several properties of the one-soliton solutions.
Indeed, as we discuss below, the mass formula for these solutions are
obtained using the same techniques as for the one-soliton solutions. In some
special cases, even n-soliton solutions of that type can be constructed.
In~\ct{acfgz} those types of solutions were constructed for the abelian  affine
Toda using the Hirota's method.

\subsection{Hirota's $\tau$-functions}

The concept of Hirota's $\tau$-function is a practical one, in the sense
that it just provides a method of constructing solutions~\ct{hiro}.
Sometimes, the definition of the $\tau$-function can be given more formally
depending upon the level of understanding of the structures of the model.
However, from the pragmatic point of view of constructing solutions, one can
say that it corresponds to a redefinition of the fields of the model such that
the equations of motion acquire a form that can be solved exactly by a {\em
formal} perturbation expansion. For some special classes of  solutions, the
perturbation procedure gives the exact solution because the expansion truncates
at a finite order. In the abelian affine Toda models, the  Hirota's
$\tau$-functions were used to construct the soliton solutions
\ct{hollo,acfgz}; but their definition was guided basically by the structure of
the equations of motion and the results known for the corresponding
one-dimensional version of that models.

An important consequence of the results obtained from the dressing
transformation method and the definition~\rf{tau} is that, now, we are able to
introduce Hirota's $\tau$-functions for any non-abelian affine Toda model.
As we will see in the examples, the Hirota's $\tau$-functions for such models
correspond to some components of the $\tau$-function $\tau_{\bf s}$ \rf{tau}
which provide a convenient parameterization of the $b$ fields. However, since
the gauge  symmetry \rf{invepml} was fixed in the definition \rf{tau}, there
will not be Hirota's $\tau$-functions associated to the gauge fixed degrees of
freedom. Roughly speaking, the Hirota's  $\tau$-functions have the form
\be
\tau_{v} = \langle v \mid \tau_{\bf s}
\lab{hirotau}
\ee
where a number of states of the representation $\mid v \rangle$ is
suitably chosen such that the $\tau_{v}$'s parameterize all the fields. In some
cases, like the principal gradation discussed in section~\ref{sec:examples},
a given $\tau_{v}$ may depend on more than one field and so it has to be split
into components by considering one parameter subspaces on the orbit defined by
$\tau_{\bf s}$. Nevertheless, it is always clear which components are needed to
describe the fields of the model.

The truncation of the Hirota's expansion occurs if, for instance, in the case
of
the one-soliton solutions, there exists a positive integer $N_v$ such
that  the generator $F$, given in \rf{epmleigen}, satisfies
\be
\langle v \mid F^{n}  \mid \l_{\bf s} \rangle = 0\, ; \qquad \qquad
\mbox{\rm for $n>N_v$}
\ee
Then, from\rf{onesol}, the corresponding Hirota's $\tau$-function for the
one-soliton solution truncate
\be
\tau_{v} = \tau^{(0)}_{v} + \tau^{(1)}_{v} + \tau^{(2)}_{v} \ldots +
\tau^{(N_{v})}_{v}
\ee
where
\be
\tau^{(n)}_{v} = {1\o n!} e^{n \gamma ( x - vt)}\langle v \mid  F^n \, \mid
\l_{\bf s} \rangle
\ee
This generalizes the case of the level-one representations of
affine simply-laced algebras where $F$ is just a nilpotent vertex operator.

\sect{The Soliton masses}
\label{sec:solmass}

\ \indent
The canonical energy-momentum tensor for the G-CAT model action
\rf{brokenwznw} is given by
\be
\Theta_{\mu\nu}= \kappa \( -{1\o 4} \Tr \( \pa_{\mu} B  \pa_{\nu} B^{-1}\) +
{1\o 8} g_{\mu\nu} \Tr\(\pa_{\rho} B  \pa^{\rho} B^{-1}\) -
g_{\mu\nu} \Tr\(\Lambda_{l} B^{-1} \Lambda_{-l}B  \) \)
\ee
and it is not traceless
\be
\Theta_{\mu}^{\mu}= -2\kappa \Tr\(\Lambda_{l} B^{-1} \Lambda_{-l}B  \)
\ee
However, when the gradation \rf{grade} is performed by a grading operator
$\qs$, $\Theta_{\mu\nu}$ can be improved by adding the term
\be
S_{\mu\nu} \equiv - {\kappa  \o 2\, l} \Tr \( \qs \(\pa_{\mu}\( B^{-1}
\pa_{\nu}
B\) - g_{\mu\nu}\pa_{\rho}\( B^{-1} \pa^{\rho} B\) \)\)
\ee
Due to the fact $B$ commutes with $\qs$, such quantity is symmetric and it is
trivially conserved
\be
\pa^{\mu} S_{\mu\nu} = 0
\ee
In addition, we have
\br
S_{\mu}^{\mu} = {\kappa  \o 2\, l} \Tr \( \qs \pa_{\mu}\( B^{-1}\pa^{\mu}B\) \)
=  2 \kappa \Tr\(\Lambda_{l} B^{-1} \Lambda_{-l}B  \)
\er
where we have used the equations of motion \rf{natoda2}, $x^{\pm} \equiv x \pm
t$, and again the fact that $B$ commutes with $\qs$.
Therefore, the improved energy-momentum tensor
\be
T_{\mu\nu} \equiv \Theta_{\mu\nu} + S_{\mu\nu}
\ee
is conserved, symmetric and traceless.

The energy-momentum tensor of the G-AT models, defined by the equations
\rf{gat}, has a simple relation with the above tensor.
Using the parameterization \rf{parb} of $B$ one has
\be
B^{-1} \pa B \mid_{\eta =0} = B_0^{-1} \pa B_0 \>+\> \partial\nu\>C
\ee
Moreover,
\br
\Theta_{\mu\nu}&=& \kappa \( -{1\o 4} \Tr \( \pa_{\mu} B_0  \pa_{\nu}
B_{0}^{-1}\) + {1\o 8} g_{\mu\nu} \Tr\(\pa_{\rho} B_{0}   \pa^{\rho}
B_{0}^{-1}\) -  g_{\mu\nu} \Tr\( e^{l\,\eta }\,\Lambda_{l} B_{0}^{-1}
\Lambda_{-l}B_{0}
\)
\)\nonu\\ &&\qquad +\kappa\( {N_{\bf s}\over 4} \( \pa_\mu \nu\> \pa_\nu \eta\>
+\>
\pa_\mu\eta\>
\pa_\nu\nu\>- \> g_{\mu\nu}\> \pa_\rho\nu\> \pa^\rho\eta\)\),
\er
and
\be
S_{\mu\nu} = -{\kappa\>N_{\bf s}\over 2l} \( \pa_\mu\pa_\nu\> \nu \>-\>
g_{\mu\nu} \pa_\rho\pa^\rho\> \nu\).
\ee
Then, one can easily verify that the
canonical energy-momentum tensor for the G-AT models is $\Theta_{\mu\nu}^{\rm
G-AT} =\Theta_{\mu\nu}\mid_{\eta =0}$ and, so,
\br
T_{\mu\nu}\mid_{\eta =0} &=&\Theta_{\mu\nu}^{\rm G-AT} +  S_{\mu\nu}\mid_{\eta
=0}\nonu\\
&=& \Theta_{\mu\nu}^{\rm G-AT}
- {\kappa  \o 2\, l} \Tr \( \qs  \(\pa_{\mu}\( \hat B_0^{-1} \pa_{\nu}
\hat{B}_0\) - g_{\mu\nu}\pa_{\rho}\( \hat{B}_0^{-1} \pa^{\rho} \hat{B}_0\)
\)\) \nonu\\
&=& \Theta_{\mu\nu}^{\rm G-AT}
-{\kappa\>N_{\bf s}\over 2l} \( \pa_\mu\pa_\nu\> \nu \>-\>
g_{\mu\nu} \pa_\rho\pa^\rho\> \nu\)
\lab{nicerel}
\er
where we have introduce the notation $\hat{B}_0 = B_0 e^{\nu\> C}$.

Now, following the reasoning of \ct{acfgz}, suppose that we have a soliton like
solution of the G-CAT model  that can be put at rest in some Lorentz frame.
Then the energy of such classical solution should be interpreted as the mass of
the soliton. But since the theory is conformally invariant, it has no mass
scale and therefore the soliton mass should vanish. If the solution is such
that the $\eta$ field vanishes then it is also a solution of the G-AT model;
this theory is not conformally invariant and, therefore, the soliton mass
evaluated with  the G-AT  energy-momentum tensor does not vanish. Using
\rf{nicerel} one observes that the contribution to the soliton mass comes from
a surface term. So, the soliton mass $M$ is then
\br
{M \, v \o \sqrt{1 - v^2}} &\equiv& \int_{-\infty}^{\infty} \, dx \,
\Theta_{10}^{\rm G-AT}\nonu\\
&=& {\kappa  \o 2\, l} \Tr \( \qs \hat{B}_0^{-1} \pa_{t} \hat{B}_0
\)\mid_{x=-\infty}^{x=\infty} = {\kappa N_{\bf s}\over 2l} \>\partial_t\>\nu
\mid_{x=-\infty}^{x=\infty}
\lab{solitonmass}
\er
where $v$ is the soliton velocity ($c=1$).

Next, using the explicit description of the integer gradations summarized in
appendix~B, we show that the trace in
\rf{solitonmass} can be expressed as an  expectation value in a given
representation.
For an integral gradation \rf{niceq}, the generators of $\hat B_0$ are the
Cartan subalgebra generators $H_a^0$, $a=1,2, \ldots r$, the central term $C$
and  the step operators $E_{\a}^n$ whose grade is zero. So, we can write
generically
\be
\hat B_0^{-1} \pa_{t} \hat B_0 = \sum_{a=1}^r y_a H_a^0 + y_0 C + \mbox{\rm
terms in the direction of $E_{\a}^n$}
\ee
{}From \rf{trace1}-\rf{trace3}
\br
\Tr \( Q_{\bf s} \hat B_0^{-1} \pa_{t} \hat B_0\)
= \sum_{a=1}^r {2 \o \a_a^2} y_a s_a + y_0 \sum_{i=0}^r s_i\> m_{i}^{\psi}.
\er
Consider now the highest weight state $\mid \l_{\bf s^{\pr}}\, \rangle$ of an
integrable representation of $\cgh$ satisfying \rf{int0}-\rf{int4}. Then we
have
\be
\langle \l_{\bf s^{\pr}} \mid   \hat B_0^{-1} \pa_{t} \hat B_0 \mid \l_{\bf
s^{\pr}}\,
\rangle = \sum_{a=1}^r y_a s^{\pr}_a + {\psi^2\o 2} y_0 \sum_{i=0}^r s^{\pr}_i
l_i^{\psi}
\ee

Therefore if we consider two different gradations ${\bf s}$ and ${\bf s^{\pr}}$
such that\footnotemark
{\footnotetext{Let us mention that these two gradations are equal, ${\bf s}=
{\bf s'}$, if either the algebra $\cg$ is simply laced, or $s_i\not=0$ only
if $\alpha_i$ is a long root of $\cg$.}}
\be
s^{\pr}_i = {\psi^2 \o \a_i^2} s_i\quad {\rm for\;\; all}\quad i=0,\ldots,r
\lab{sprime}
\ee
then one gets
\be
{M \, v \o \sqrt{1 - v^2}} = {\kappa  \o \psi^2\, l} \langle \l_{\bf s^{\pr}}
\mid  \hat B_0^{-1} \pa_{t}\> \hat B_0 \mid \l_{\bf s^{\pr}}\, \rangle
\mid_{x=-\infty}^{x=\infty}
\ee
Notice that the subalgebras $\cgh_0$ and $\cgh_0^{\pr}$ of zero grade with
respect to the gradations ${\bf s}$ and ${\bf s^{\pr}}$, respectively, are
equal. Therefore, $\mid \l_{\bf s^{\pr}}\, \rangle$ is an eigenstate of $B_0$,
and so
\be
{M \, v \o \sqrt{1 - v^2}} = -{\kappa\o \psi^2\, l} \pa_t \ln  \langle \l_{\bf
s^{\pr}}
\mid  \hat B_0^{-1} \mid \l_{\bf s^{\pr}}\, \rangle \mid_{x=-\infty}^{x=\infty}
\ee

{}From \rf{bbar} and \rf{b0vac} one has, $\pa_t \ln  \langle \l_{\bf s^{\pr}}
\mid  \hat B_0^{-1} \mid \l_{\bf s^{\pr}}\, \rangle
\mid_{x=-\infty}^{x=\infty}=
\pa_t \ln  \langle \l_{\bf s^{\pr}} \mid  b^{-1} \mid \l_{\bf s^{\pr}}\,
\rangle \mid_{x=-\infty}^{x=\infty}$. As we have seen in \rf{gsol}, the
soliton solutions correspond, in \rf{important}, to the choices $h_L={\tilde
h}_R=1$ and $g=e^F$ with $F$ being an eigenvector of $E_{\pm l}$. So, from
\rf{important}, \rf{epmleigen} and \rf{solpar}, it follows that
\be
M = \kappa \, {\gamma \sqrt{1-v^2}\o \psi^2 l}\,\,{{\langle \l_{\bf s^{\pr}}
\mid  e^{\gamma (x-vt)} F \exp (e^{\gamma (x-vt)} F) \mid \l_{\bf s^{\pr}}\,
\rangle}\o{\langle \l_{\bf s^{\pr}}
\mid  \exp (e^{\gamma (x-vt)} F) \mid \l_{\bf s^{\pr}}\,
\rangle}}\mid_{x=-\infty}^{x=\infty}
\lab{solitonmass0}
\ee
If there exists a positive integer $N^{\pr}_{\rm max.}$ such that $F$
satisfies
\be
\langle\,\l_{\bf s^{\pr}}\,\mid F^{N^{\pr}_{\rm max.}} \mid \l_{\bf s^{\pr}}\,
\rangle \neq 0 \, ; \qquad \qquad \mbox {\rm for $ N^{\pr}_{\rm max.}>0$}
\ee
and also
\be
\langle\,\l_{\bf s^{\pr}}\,\mid F^{n+1} \mid \l_{\bf s^{\pr}}\,
\rangle =0 \, ; \qquad \qquad \mbox {\rm for $n\geq N^{\pr}_{\rm max.}$}
\ee
then, for $\gamma >0$ only the upper limit in \rf{solitonmass0} contributes,
and if $\gamma <0$ only the lower one, so
\be
M = \kappa \,{N^{\pr}_{\rm max.}\o \psi^2 l}\,\mid\gamma \mid\sqrt{1-v^2}  =
\kappa \, {2 \o \psi^2}\, {N^{\pr}_{\rm max.}\o l}\, \sqrt{-\omega_l \,
\omega_{-l}}
\lab{solitonmass2}
\ee
where we used \rf{solpar}.

Eq. \rf{solitonmass2} generalizes to the G-CAT models the mass formula
obtained in \ct{acfgz,otu} for the solitons in the abelian  Affine Toda models.
However, a detailed analysis of each particular G-CAT model has to be done to
ensure the physical significance of \rf{solitonmass2}. The basic
point of that analysis is to establish that the eigenvalues of $E_{\pm l}$,
namely $\omega_{\pm l}$, are such that $\gamma$ and $v$ are real.

Notice that if we had taken in \rf{vacgp}, $\eta= \eta_0 = {\rm constant}$,
then $E_l$ would get multiplied by the factor $v_{\eta}$, introduced in
\rf{higgs}. Therefore the eigenvalue $\omega_l$, and consequently the soliton
masses, would also get multiplied by the same factor. This is an evidence that
the Higgs like mechanism discussed at the end of section~\ref{sec:classmass} is
also working in the generation of the soliton masses.

The mass formulas for the fundamental particles \rf{partmass} and for the
one-soliton solutions~\rf{solitonmass2} indicate a very deep structure in such
models, which is still to be understood. They might indicate a sort of
duality similar to the electromagnetic duality conjectured by Montonen and
Olive
\ct{duality} in four dimensional gauge theories involving the interplay of
monopoles and gauge particles. In fact, the use of the gradation ${\bf
s^{\pr}}$ defined in \rf{sprime}, which is a technical artifact here to
calculate the soliton masses, is an indication of the interplay between the
algebra $\cgh$ and its dual $\cgh^{v}$ ---roots interchanged by co-roots---
similar to the electromagnetic duality. Some results concerning duality in
Toda models were recently investigated in~\ct{watts,grisaru}.

Since $F$ is an eigenvector of $E_{\pm l}$ then it is obviously an eigenvector
of $\lb E_{-l}\, , \, \lb E_{l} \, , \, \cdot \rb \rb$. So, expanding it in
eigenvectors of the grading operator $Q_{\bf s}$
\be
F = \sum_{n} F^{(n)} \, ; \qquad \qquad
\lb Q_{\bf s} \, , \, F^{(n)} \rb = n F^{(n)}
\ee
we observe that each component satisfies
\be
\lb E_{-l}\, , \, \lb E_{l} \, , \, F^{(n)} \rb \rb = \omega_l
\omega_{-l} F^{(n)}
\ee
Therefore if the zero mode $F^{(0)}$ is non-vanishing, we have a generator of
$\cgh_0$ with eingenvalue $\l = \omega_l \omega_{-l}$ and so, from
\rf{partmass} a fundamental particle of mass $m^2 = -4 \omega_l\omega_{-l}$.
So, for each $F$ with non-vanishing zero mode we can put in correspondence a
soliton and a fundamental particle with the masses satisfying
\be
m_{\rm part.} = {l\o \kappa } \, {\psi^2\o N^{\pr}_{\rm max.} } \,
M_{\rm sol.}
\lab{solpartmass}
\ee
Obviously any two masses can be related by a proportionality constant.
However, in the case of the abelian affine Toda models \ct{otu,acfgz}, the
integers $N^{\pr}_{\rm max.}$ for each species of particles and solitons, are
such that the particle and soliton masses are proportional to the right and
left Perron-Frobenius vectors respectively. This is what indicates the
existence of a duality symmetry in the abelian affine Toda models. There
remains to establish the types of non-abelian affine Toda models that present a
similar relation. In any case, it is a remarkable fact that the soliton and
particle masses are both determined by the eigenvalues of $E_{\pm l}$ in any
one of these models.

\sect{Examples}
\label{sec:examples}

\ \indent
Here, we work out in detail some aspects of the G-CAT models associated
to the principal and to the homogeneous gradation of $\cgh$, in which the
dimension of $\cgh_0$ is minimal and maximal, respectively. For the
principal gradation, we recover the abelian CAT model of~\ct{AFGZ,bb}, and we
show that the Hirota's $\tau$-functions defined in~\ct{acfgz,hollo} easily
follow from the systematic approach of section~\ref{sec:tau}. For the
homogeneous gradation, we construct the one-soliton solutions when the finite
Lie algebra $\cg$ is either compact or non-compact; in the former case, the
resulting model has very suggestive features.

\subsection{Principal gradation}

\ \indent
The principal gradation of $\cgh$ is defined by $s_i =1$, $i=0,1,2 \ldots r$
(see \rf{gradop}). The highest weight state in the definition of the
$\tau$-function
\rf{tau} is given by \rf{lambdas}, {\it i.e.\/},
\be
\mid \l_{\rm ppal} \rangle \equiv \bigotimes_{i=0}^r \mid {\hat
\l}_{i}\, \rangle
\ee

{}From \rf{int1} we get
\be
h_i \, \mid \l_{\rm ppal} \rangle = \mid \l_{\rm ppal} \rangle \, ; \qquad
\qquad i=0,1, \ldots r
\ee

For the principal gradation the subalgebra $\cgh_0$ is generated by $h_i$ and
$D$. So, we parameterize the group element $b$, in \rf{tau} as
\be
b = \exp \( \sum_{i=0}^r \vp^i \, h_i \)
\lab{bppal}
\ee
In this case the zero mode $\tau^{(0)}_{\rm ppal}$ \rf{modeexp} of the
$\tau$-function depends on all $b$ fields. From \rf{tau0} and \rf{bppal}
\be
\tau^{(0)}_{\rm ppal} =  b^{-1} \mid \l_{\rm ppal} \rangle =
e^{-\sum_{i=0}^r\vp^i}\, \mid \l_{\rm ppal} \rangle
\ee
We then define the Hirota's $\tau$-functions, in this case, as
\be
\tau_i \equiv \langle  \l_{\rm ppal} \mid e^{-\vp^i h_i}\, \mid \l_{\rm ppal}
\rangle =  \langle  {\hat \l}_{i} \mid  b^{-1} \, \mid {\hat \l}_{i}
\rangle = e^{-\vp^i}
\ee
and so
\be
\vp^i = - \ln \tau_i \,  \qquad \qquad  i=0,1,2, \ldots r
\ee
Therefore, the Hirota's $\tau$-functions correspond to the components, in the
tensor product space of the representation, of $\tau^{(0)}_{\rm ppal}$. If we
had parameterized $b$ as
\be
b=\exp \(\sum_{a=1}^r\phi^a H_a^0 + \nu C\)
\lab{bppal2}
\ee
we would get, since $\nu = {2\o \psi^2}\vp^0$ and $\phi^a = \vp^a - l_a^{\psi}
\vp^0$, that
\be
\nu = - {2\o \psi^2} \ln \tau_0
\, ; \qquad
\phi^a = - \ln {\tau_a \o (\tau_0)^{l_a^{\psi}}}
\, ; \qquad a=1,2, \ldots r
\ee
This is exactly the definition of $\tau$-function used in~\ct{acfgz} to
construct the soliton solutions of the abelian affine Toda models by the
Hirota method (except for some terms associated to the fact that,
in~\ct{acfgz}, the element $b_0 e^{\nu_0  C}$ was not factorized as in
\rf{bbar}).

For the case $l=1$ one can take $\Lambda_{\pm l}$ in \rf{const1}-\rf{const2}
to be
\br
\Lambda_{1} = \sum_{i=0}^r {\bar q}_i e_i \, ; \qquad  \qquad
\Lambda_{-1} = \sum_{i=0}^r q_i f_i
\er
for some constants $q_i$ and ${\bar q}_i$. The element $b_0$ in \rf{epml} can
be taken as
\be
b_0 = e^{\gamma \cdot H^0 + \rho \, D} \, ; \qquad \mbox{\rm such that} \,
\qquad b_0 e_i b_0^{-1}  \equiv {l_i^{\psi}\o {q_i
{\bar q}_i}} e_i
\ee
with $l_a^{\psi}$  given in \rf{lipsi}, and so
\be
E_{1} = b_0 \Lambda_{1} b_0^{-1} = \sum_{i=0}^r {l_i^{\psi}\o {q_i}} e_i
\, ; \qquad \qquad E_{-1} =\Lambda_{-1}
\ee
satisfies
\be
\lb E_1 \, , \, E_{-1} \rb = {2\o {\psi^2}}\, C
\ee
The symmetry groups given in \rf{invepml}, namely $h_L$ and ${\tilde h}_R$, are
trivial in this case, {\it i.e.\/}, they are exponentiations of the central
term $C$. On the other hand, in the purely imaginary coupling constant
regime, the transformations \rf{degenerate} are responsible for the infinitely
degenerate vacua leading to the topological soliton solutions.

Therefore from \rf{gcatnice} one gets the equations of motion for the
abelian Conformal Affine Toda (CAT) models \ct{AFGZ}
\br
\pa_{+} \pa_{-} \phi^a &=& - l_a^{\psi}\( e^{K_{ab}\, \phi^b} - e^{K_{0b}\,
\phi^b}\) e^{\eta}\\
\pa_{+} \pa_{-} \eta &=& 0\\
\pa_{+} \pa_{-} \nu &=& {2\o {\psi^2}}\(1 -e^{K_{0b}\phi^b}\,  e^{\eta} \,
\)
\er
and so $\phi^a=\eta =\nu =0$ is a vacuum solution.

The soliton solutions and integrability properties of those models were
extensively discussed in the literature \ct{AFGZ,acfgz,otu,charges,freeman2}
and we do not describe them here.

If one takes $l>1$ one obtains different models. However, for the principal
gradation, and for any gradation where all $s_i$'s are non-vanishing, the
subalgebra $\cgh_0$  is generated by $h_i$, $i=0,1,2 \ldots r$, and
$Q_{\bf s}$. Therefore, the generators of grade $l$ are step operators
which corresponding opposite root step operators have grade $-l$. Then,
the difference of two roots of grade $l$ is never a root. In this sense, the
roots of grade $l$ behave like  a set of simple roots for some subalgebra of
$\cgh$, not necessarily simple. So, this indicates that by considering models
with $l>1$ for gradations where all $s_i$'s are non-vanishing one is going to
get abelian Toda models (affine or not) for different subalgebras of $\cgh$.

\subsection{Homogeneous Gradation}

In this case all $s_i$'s vanish except $s_0$ which equals one. So, the grading
operator is just $D$
\be
Q_{\rm hom.} \equiv D
\ee
The subalgebra $\cgh_0$ is generated by $H_a^0$, $a=1,2, \ldots r$, $E_{\pm
\a}^0$, $D$ and $C$, {\it i.e.\/}, it is the simple algebra $\cg$ in addition
to $D$ and $C$. There is a great variety of choices for $E_{\pm l}$, but here
we will consider those that can be rotated into the $H^{\pm l}$ subspace. So,
we take
\be
E_{l} = \sum_{a=1}^r q_a \, H_a^l \equiv q \cdot H^l\, ; \qquad \qquad
E_{-l} = \sum_{a=1}^r {\bar q}_a \, H_a^{-l} \equiv {\bar q} \cdot H^{-l}
\lab{homepml}
\ee
Therefore from \rf{km1} and \rf{vacua} we have
\be
\b = l \sum_{a,b=1}^r  q_a \eta_{ab} {\bar q}_b \equiv l \, q \cdot {\bar q}
\ee

The eigensubspaces $\cgh_n$ of $Q_{\rm hom.}$ are all copies of the adjoint
representation of $\cgh_0$, and therefore one does not expect the properties
of the model to change by varying $l$. For instance, for any $l$, the symmetry
groups of the model given in \rf{invepml}, are the same and equal to the Cartan
subgroup of $\cgh_0$, {\it i.e.\/},
\be
h_L \, \, \, {\rm and} \, \, \, {\tilde h}_R \, = \, \mbox{\rm exponentiations
of $H_a^0$  for $a=1,2, \ldots r$};
\lab{gaugehom}
\ee
this in agreement with the comments below eq.~\rf{brokenwznwplus}.
Notice that if $q$ and/or ${\bar q}$ are orthogonal to some root $\a$, then
the  corresponding step operators $E_{\pm \a}^0$ are also generators of
${\tilde h}_R$ and/or $h_L$.  According to the discussion of
section~\ref{sec:classmass} the model has at least rank-$\cg$$+1$ massless
particles (depending upon the choice of $q$ and ${\bar q}$). In fact, the
diagonalization of the operator in \rf{masseigen} is quite simple (see
\rf{km1}-\rf{km6})
\br
\lb E_{-l} \, , \lb E_{l}\, , \, C \rb\rb &=& 0\\
\lb E_{-l} \, , \lb E_{l}\, , \, H_a^0 \rb\rb &=& 0\\
\lb E_{-l} \, , \lb E_{l}\, , \, E_{\pm \a}^0 \rb\rb &=& (q\cdot \a ) \, ({\bar
q}\cdot \a ) \, E_{\pm \a}^0
\er
So, there is a mass degeneracy associated to each pair of roots $\pm \a$.
In order to have non-negative masses one has, from \rf{partmass}, to take both
$q$ and $(-{\bar q})$ (or equivalently $(-q)$ and ${\bar q})$) lying in the
Fundamental Weyl chamber \ct{hump}. However, such statement is misleading.
Going back to the Klein-Gordon equation~\rf{kge}, one notices that the trace
form multiplies the mass and the kinetic terms of the corresponding
Lagrangian. Therefore if the trace form restricted to $\cgh_0$ (more
properly, to $\cgh_{0}^{\ast}$) is not positive definite, we have a model where
the energy is not bounded below. However, as we pointed out in
sections~\ref{sec:reduction} and~\ref{sec:symmetries}, we can always choose the
real form of the  Kac-Moody algebra $\cgh$ such that the trace form restricted
to $\cgh_0$ is positive definite. In the case of the homogeneous
gradation we can take
$\cgh_0$ to be generated by $H_a^0$, $E_{\a}^0 + E_{-\a}^0$, and $i\( E_{\a}^0
-
E_{-\a}^0\)$, in addition to $C$ and $D$. So, from \rf{trace1}-\rf{trace3} the
trace form is positive definite except  for the subspace generated by $C$ and
$D$. However, since we are considering the  model with the conformal symmetry
spontaneously broken by $\eta = \eta_0 =  {\rm const.}$ (see \rf{higgs}), one
gets that the field in the direction of  the central term does not contribute
to the energy, since it is orthogonal to  itself and to all remaining
generators, see~\rf{brokenwznwplus}.  Therefore we have a theory with masses
and
energies real and non-negative. We have rank-$\cg$$+1$ massless particles and a
number of massive particles equal to twice the number of positive roots and
masses given by (see
\rf{partmass})
\be
m^2_{\a} = - 4\, (q\cdot \a ) \, ({\bar q}\cdot \a ) \, ; \qquad
\qquad \mbox{\rm for any positive root $\a$ of $\cg$}
\lab{hompartmass}
\ee
with $q$ and $(-{\bar q})$  being arbitrary vectors lying in the Fundamental
Weyl chamber.

The highest weight state $\mid \l_{\bf s} \rangle$ in this case is given by
(see \rf{lambdas})
\be
\mid \l_{\rm hom.} \rangle \equiv \mid {\hat \l}_{0} \rangle
\ee
with ${\hat \l}_{0}$ given by \rf{lambda0}. So, we have the scalar
representation of $\cgh_0$.

We parameterize the group element $b$ introduced in \rf{bbar} as
\be
b = {\tilde b}\, e^{{2\o\psi^2}\nu \, C}
\lab{btilde}
\ee
Then, from \rf{tauhat0} and \rf{fundrep0} we get
\be
{\hat \tau}_{\rm hom.}^{(0)} = e^{-\nu}\, ; \qquad {\rm or} \qquad
{\tau}_{\rm hom.}^{(0)} =  e^{-\nu} \mid {\hat \l}_{0} \rangle
\ee
Therefore, ${\tau}_{\rm hom.}^{(0)}$ is not sufficient to parameterize the
$b$-fields. In order to do that we need to consider more components of the
$\tau$-function.

We write the quantity $t^{(-l)}$ defined in \rf{tpms} as
\be
t^{(-l)} \equiv t^{(-l)}_H \cdot H^{-l} + \h \sum_{\a >0} {\a^2\o 2}
\( t^{(-l)}_{\a ,(1)} ( E_{\a}^{-l} + E_{-\a}^{-l}) +
t^{(-l)}_{\a ,(2)} ( E_{\a}^{-l} - E_{-\a}^{-l})/i \)
\lab{tml}
\ee
Substituting that in \rf{tml2}, and using \rf{homepml}, \rf{btilde} and
\rf{trace1}-\rf{trace3} we get (fixing the gauge by ${\tilde h}_R =1$)
\br
{2\o\psi^2}\,\pa_{+} \nu  &=& l\, q \cdot t^{(-l)}_H
\lab{tmlhom1}\\
\Tr \( {\tilde b}^{-1} \pa_{+} {\tilde b}\, H_a^0\) &=& 0 \, ; \qquad\qquad
a=1,2,\ldots r
\lab{tmlhom2}\\
\Tr \( {\tilde b}^{-1} \pa_{+} {\tilde b}\, ( E_{\a}^{0} + E_{-\a}^{0})\) &=&
-i \, q \cdot \a \, t^{(-l)}_{\a ,(2)}
\lab{tmlhom3}\\
\Tr \( {\tilde b}^{-1} \pa_{+} {\tilde b}\, ( E_{\a}^{0} - E_{-\a}^{0})/i\) &=&
i \, q \cdot \a \, t^{(-l)}_{\a ,(1)}
\lab{tmlhom4}
\er

We shall restrict ourselves to the solutions of the type described in
\rf{easier}. Therefore, from~\rf{modeexp} and~\rf{nuestra},
\be
\tau^{(-s)}_{\rm hom} = 0 \, ; \qquad 0<s<l
\lab{nicechoice}
\ee
and
\be
\tau^{(-l)}_{\rm hom} = -e^{-\nu} t^{(-l)} \, \mid  {\hat \l}_0\rangle
\ee

Following \rf{hirotau}, we then define the Hirota $\tau$-functions as
\br
\tau_0 &\equiv& \langle {\hat \l}_0 \mid \tau_{\rm hom} = e^{-\nu}
\lab{hirohom1}\\
\tau_{\a , (1)} &\equiv& \langle {\hat \l}_0 \mid ( E_{\a}^{l} +
E_{-\a}^{l})\tau_{\rm hom} = -l\, {\psi^2\o 2}\, t^{(-l)}_{\a ,(1)}\, \tau_0
\lab{hirohom2}\\
\tau_{\a , (2)} &\equiv& \langle {\hat \l}_0 \mid (( E_{\a}^{l} -
E_{-\a}^{l})/i)\tau_{\rm hom} = -l\, {\psi^2\o 2}\, t^{(-l)}_{\a ,(2)}\, \tau_0
\lab{hirohom3}
\er
for every positive root $\a$.

Notice there are no Hirota $\tau$-functions associated to the $b$-fields in
the direction of the Cartan subalgebra generators $H_a^0$. The reason is that
we have gauge fixed the symmetry of the model by setting in \rf{gaugehom},
$h_L={\tilde h}_R=1$. Indeed from \rf{tmlhom2}, we see that ${\tilde b}^{-1}
\pa_{+} {\tilde b}$ has no component in the direction of the Cartan
subalgebra.

The relation between the Hirota $\tau$-functions and the fields of the model
can be extracted from \rf{hirohom1}-\rf{hirohom3} and
\rf{tmlhom1}-\rf{tmlhom4}. In this case, such relations involve derivatives of
the fields. The Hirota equations can them be obtained from the equations of
motion for the fields. However we do not have to solve such Hirota equations,
because the solutions have already been constructed by the dressing
transformation method. In order to obtain the soliton solutions, we choose the
constant group element $g$ in \rf{important} as in \rf{gsol}.

\subsubsection{Non-compact solitons}

As we have seen in section~\ref{sec:tau} the one-soliton solutions are
associated to the eigenvectors of $E_{\pm l}$. In this case we have that
\br
F_a^k (\zeta ) &=& \sum_{n \in \IZ} \zeta^{-nl} H_a^{k + nl}\, ;
\qquad a=1,2, \dots r \, ; \qquad k=1,2, \dots l-1
\lab{f1}\\
F_{\pm\a}^p (\zeta ) &=& \sum_{n \in \IZ} \zeta^{-nl} E_{\pm\a}^{p + nl}
\, ; \qquad p=0,1,2, \dots l-1
\lab{f2}
\er
are eigenvectors of $E_{\pm l}$, given in \rf{homepml}
\br
\lb E_{\pm l} \, , \, F_a^k (\zeta ) \rb &=& 0 \, ; \qquad \mbox{\rm for $k\neq
0$}
\lab{eigenhom1}\\
\lb E_{l} \, , \, F_{\pm\a}^p (\zeta ) \rb &=& \pm \zeta^{l}(q\cdot \a )
F_{\pm\a}^p (\zeta )
\lab{eigenhom2}\\
\lb E_{-l} \, , \, F_{\pm\a}^p (\zeta ) \rb  &=& \pm \zeta^{-l}({\bar q}\cdot
\a
) F_{\pm\a}^p (\zeta )
\lab{eigenhom3}
\er
where $\zeta$ is a complex parameter.

Notice that if we take $k=0$ in \rf{f1} we do not get zero eigenvectors, since
\be
\lb E_{l} \, , \, F_a^0 (\zeta ) \rb = l\,\zeta^l{2 \a_a \cdot q\o \a_a^2} \, C
\, ; \qquad
\lb E_{-l} \, , \, F_a^0 (\zeta ) \rb = -l\,\zeta^{-l}{2 \a_a \cdot {\bar q}\o
\a_a^2} \, C
\ee
One could choose $q$ and ${\bar q}$ to be orthogonal to all simple roots $\a_a$
except one, by taking them to lie in the direction of one of the fundamental
weights. Therefore one could have rank-$\cg$$-1$ eigenvectors of type $F_a^0$
with zero eigenvalue.

However the $F$'s with zero eigenvalue lead to  constant solutions in
\rf{onesol}, since from \rf{solpar} we get that $\gamma = 0$. Therefore,
there are no one-soliton solutions associated to $F_a^k$ given in \rf{f1}.
As we have commented on the paragraph below \rf{easier}, we can not take $F$ in
\rf{gsol} as one of the $F_{\pm\a}^p$ for $p \neq 0$, given in \rf{f2}, because
\rf{nicechoice} will not be satisfied. So, we will construct for each
$F_{\pm\a}^0$, a one-soliton solution.

Therefore from \rf{onesol}, \rf{solpar}, \rf{eigenhom2}, \rf{eigenhom3} and
\rf{hirohom1}-\rf{hirohom3} we get two one-soliton solutions for each
positive root $\a$, with the Hirota's $\tau$-functions given by
\br
\tau_0^{1-sol,\a} &=& 1
\lab{1solalpha1}\\
\tau_{\a , (1)}^{1-sol,\a} &=& l\, \zeta^l {\psi^2\o \a^2}\, e^{\gamma_{\a}
(x-v_{\a}t)}
\lab{1solalpha2}\\
\tau_{\a , (2)}^{1-sol,\a} &=& i\, l\, \zeta^l {\psi^2\o \a^2}\, e^{\gamma_{\a}
(x-v_{\a}t)}
\lab{1solalpha3}\\
\tau_{\b , (1)}^{1-sol,\a} &=&\tau_{\b , (2)}^{1-sol,\a} = 0 \, ; \qquad
\mbox{\rm for $\b \neq \a$}
\lab{1solalpha4}
\er
for $g$ in \rf{gsol}, given by
\be
g_{1-sol,\a}(\zeta ) \equiv e^{F_{\a}^0(\zeta )}
\ee
and the parameters of the solitons being
\be
\gamma_{\a}(\zeta ) = (q\zeta^l -{\bar q}\zeta^{-l})\cdot \a \, ; \qquad
 v_{\a}(\zeta ) = -{{(q\zeta^l+{\bar q}\zeta^{-l})\cdot \a}\o{(q\zeta^l-{\bar
q}\zeta^{-l})\cdot \a }}
\lab{paralfa1}
\ee
Notice that, for $\zeta$ real, we have $\mid v_{\a} \mid <1$, since $(q\cdot \a
)({\bar q}\cdot \a )<0$ for $\a >0$ ($q$ and $(-{\bar q})$ belong to the
Fundamental Weyl chamber).
The corresponding solution for the $b$-fields can be obtained from
\rf{tmlhom1}-\rf{tmlhom4} and \rf{hirohom1}-\rf{hirohom3}, giving ($\nu =0$,
see \rf{btilde})
\be
{\tilde b}_{1-sol,\a} =   \exp \( - e^{\gamma_{\a}(x-v_{\a}t)}\,
E_{\a}^0\)
\lab{solution1}
\ee
Therefore, such one-soliton solution leads to an element of the non-compact
real form of the group, as it ``excites'' only the field in the direction
of $E_{\a}^0$.

In addition, for $g$ in \rf{gsol}, given by
\be
g_{1-sol,-\a}(\zeta ) \equiv e^{F_{-\a}^0(\zeta )}
\ee
we have
\br
\tau_0^{1-sol,-\a} &=& 1
\lab{1solmalpha1}\\
\tau_{\a , (1)}^{1-sol,-\a} &=& l\,\zeta^l\, {\psi^2\o \a^2}\, e^{\gamma_{-\a}
(x-v_{-\a}t)}
\lab{1solmalpha2}\\
\tau_{\a , (2)}^{1-sol,-\a} &=& -i\, l\,\zeta^l\, {\psi^2\o \a^2}\,
e^{\gamma_{-\a} (x-v_{-\a}t)}
\lab{1solmalpha3}\\
\tau_{\b , (1)}^{1-sol,-\a} &=&\tau_{\b , (2)}^{1-sol,-\a} = 0 \, ; \qquad
\mbox{\rm for $\b \neq \a$}
\lab{1solmalpha4}
\er
where
\be
\gamma_{-\a}(\zeta ) = -(q\zeta^l -{\bar q}\zeta^{-l})\cdot \a \, ; \qquad
 v_{-\a}(\zeta ) = -{{(q\zeta^l+{\bar q}\zeta^{-l})\cdot \a}\o{(q\zeta^l-{\bar
q}\zeta^{-l})\cdot \a }}
\lab{paralfa2}
\ee
Again, we have $\mid v_{-\a} \mid <1$ for $\zeta$ real and  $(q\cdot\a )({\bar
q}\cdot \a )<0$. Notice that for a given $\zeta$, the one-soliton solutions
associated to $F_{\a}^0(\zeta )$ and $F_{-\a}^0(\zeta )$  have the same
velocity
but opposite width parameter $\gamma$ (see \rf{paralfa1} and \rf{paralfa2}).
The  corresponding solution for the $b$-fields is similarly obtained from
\rf{tmlhom1}-\rf{tmlhom4} and \rf{hirohom1}-\rf{hirohom3}, giving ($\nu =0$,
see \rf{btilde})
\be
{\tilde b}_{1-sol,-\a} =   \exp \( - e^{\gamma_{-\a}(x-v_{-\a}t)}\,
E_{-\a}^0\)
\lab{solution2}
\ee
One can easily verify that \rf{solution1} and \rf{solution2} are indeed
solutions, by substituting them directly into \rf{gcatnice}.

All these one-soliton solutions have vanishing masses. The reason is that, from
\rf{solitonmass}, one observes that, only the component of $B_0^{-1} \pa_t B_0$
in the direction of the central term $C$ contributes to the mass,  since
$Q\equiv D$ in this case. But, from \rf{bbar}, \rf{btilde}
\rf{b0vac}, \rf{1solalpha1} and \rf{1solmalpha1} that contribution is
$\pa_t (\nu_0 + {2\o \psi^2} \nu)C \sim t \, C$, and so, $\Tr \( Q \, B_0^{-1}
\pa_t B_0\) \mid_{x=-\infty}^{x=\infty}=0$. Another way of getting that is to
use \rf{solitonmass0} and the fact that the expectation value  of any
non-vanishing positive power of $F_{\pm \a}^0(\zeta )$, in the highest weight
state
$\mid {\hat \l}_0 \rangle$, is zero. Therefore, although we can put the
one-soliton solutions in one to one correspondence with the massive
fundamental particles, through the eigenvalues of $E_{\pm l}$, we do not have
in this case, the proportionality of the masses as discussed in
\rf{solpartmass} ($N^{\pr}_{\rm max.}=0$). In the G-CAT model we do have
massless solitons travelling with velocity $\mid v \mid <1$. That
is because the conformal symmetry does not allow a mass scale. It would be
interesting to understand from a physical point of view why the above
solitons are massless.

As we have said, the one-soliton solutions associated to the eigenvectors
$F_{\pm\a}^{p+nl}$ for $p\neq 0$ can not be obtained by the above procedure due
to the choice \rf{nicechoice} we have made (see \rf{easier}). However, one does
not expect them to correspond to new soliton solutions. The reason is that
they do not exist in the case $l=1$, and the equations of motion do not change
considerably by changing $l$, as commented below eq.\rf{brokenwznwplus}. As we
pointed out above, the subspaces
$\cgh_n$ are copies of the adjoint representation of $\cgh_0$ and the only
equation of motion dependent on $l$ (for $\eta =0$) is that for the $\nu$
field. In addition, the  eigenvectors $F_{\pm\a}^{p+nl}$ can  be obtained from
$F_{\pm \a}^{nl}$ by
\be
F_{\pm \a}^{p+nl} = \pm \lb {\a \cdot H^p\o \a^2} \, , \, F_{\pm \a}^{nl} \rb
\, ; \qquad p=1,2,\ldots l-1
\ee
and ${\a \cdot H^p\o \a^2}$ lies in the centralizer of $E_{\pm l}$. This is in
fact some sort of symmetry of the model. If $T$ is an eigenvector of $E_{\pm
l}$, $\lb E_{\pm l}\, , \, T \rb = \l_{\pm} T$, and if $L$ lies in the
centralizer of $E_{\pm l}$, then $\lb L \, , \, T\rb$ is also an eigenvector
of $E_{\pm l}$ with the same eigenvalue.

\subsubsection{Compact solitons}

The one-soliton solutions for the compact real form of the group can
be directly constructed in terms of the fields. Introducing the
generators
\be
T_1 (\a ) \equiv \h \( E_{\a} + E_{-\a}\) \, ; \qquad
T_2 (\a ) \equiv {1\o 2\, i} \( E_{\a} - E_{-\a}\),
\ee
one can easily verify that the group elements
\be
b^{(j)}\equiv e^{i\, \vp\, T_j (\a )}\, e^{\nu \, C}\, ; \qquad
j=1,2
\lab{compsol}
\ee
are solutions of the equations of motion \rf{gcatnice}, for $\eta =0$ and
$E_{\pm l}$ given by \rf{homepml}, if the fields $\vp$ and $\nu$ satisfy
($\pa_{+}\pa_{-}=-{1\o 4} \pa^2 =-{1\o 4}(\pa_t^2 - \pa_x^2)$)
\br
\pa^2 \vp &=& - m^2_{\a} \sin \vp
\lab{sg1}\\
\pa^2 \nu &=& -{l\o \a^2}\, m^2_{\a}\( \cos \vp - 1\)
\lab{sg2}
\er
where $m^2_{\a}$ is given in \rf{hompartmass}.

Therefore, for every solution of the Sine-Gordon model we get two
solutions, corresponding to $b^{(1)}$ and $b^{2)}$ given in \rf{compsol}, of
the G-AT model associated to the homogeneous gradation. The
structure of these models is such that each $SU(2)$
subgroup associated to a positive root $\a$ can be considered independently.

We then have topological solitons, and the infinitely many vacua are
associated to the invariance of the model under the transformations
\be
b \ra b \, e^{2 \pi i\, n\, T_j(\a )}\, ; \qquad
b \ra e^{2 \pi i\, n\, T_j(\a )}\, b
\lab{real2}
\ee
since the generators \rf{homepml} satisfy
\be
e^{2 \pi i\, n\, T_j(\a )}\, E_{\pm l} \, e^{-2 \pi i\, n\, T_j(\a )} =
E_{\pm l}
\ee
Since we are working with the compact real form and $b$ is parameterized as in
\rf{bbar2}, \rf{real1} implies a real transformation on the fields of the
model.

The operators \rf{homepml} are also certainly invariant under the
transformations \rf{degenerate}, which imply the fields transform as
\be
b \ra e^{2\pi i(\mu^v \cdot H^0 + n \, D)} \, b \, ; \qquad
b \ra   b \, e^{2\pi i(\mu^v \cdot H^0 + n \, D)}
\lab{real1}
\ee
with $\mu^v$ a co-weight and $n$ and integer. However, these are not relevant
here, because the solutions we are considering do not possess fields in the
direction of the Cartan subalgebra (except for $\nu$); those have been gauge
fixed by setting $h_L={\tilde h}_R =1$ in \rf{gaugehom}.

In order to evaluate the masses of the solutions \rf{compsol} which can be put
at rest in some Lorentz frame, we use the formula \rf{solitonmass}. Since the
grading operator in this case is just $D$, and that is orthogonal to all
generators except $C$, we get (see \rf{bbar})
\be
{M \, v \o \sqrt{1 - v^2}} = {\kappa  \o 2\, l} \pa_t \( \nu + \nu_0 \)
\mid_{x=-\infty}^{x=\infty} = {\kappa  \o 2\, l} \pa_t \, \nu
\mid_{x=-\infty}^{x=\infty}
\lab{massaux}
\ee
where we have used the fact that $\nu_0 \sim x^2 - t^2$ (see \rf{b0vac}).

The one-soliton solution of the Sine-Gordon model is
\be
\vp_{\rm 1-sol} = - 4 \epsilon_0 \arctan \exp \(
{m_{\a}(x-vt-x_0)\o{\sqrt{1-v^2}}}\)
\lab{onesolsg}
\ee
where $\epsilon_0$ is the topological charge and $x_0$ the initial position of
the soliton. For the fundamental solitons, where $\epsilon_0 = \pm 1$, one
gets from \rf{sg2}, \rf{massaux} and \rf{onesolsg} that the soliton masses are
given by
\be
M_{\a}^{\rm sol.} = \kappa \, {2\o \a^2}\, m_{\a}^{\rm part.}
\lab{solpartmasshom}
\ee
where $m_{\a}^{\rm part.}$ is given in \rf{hompartmass}, and $\kappa$ is the
overall factor in the action \rf{brokenwznw}.

Therefore for each positive root $\a$ we have two fundamental particles with
the same mass  $m_{\a}^{\rm part.}$, and  two one-soliton solutions,
corresponding to $b^{(1)}$ and  $b^{(2)}$ given by \rf{compsol} and $\vp$
being the one-soliton of the S-G model, with equal masses $M_{\a}^{\rm sol.}$.
The relation between soliton and particle masses is given simply by
\rf{solpartmasshom}. That is a remarkable relation. Like in the abelian
affine Toda models, it indicates the existence of a duality transformation that
interchanges the roles of particles and solitons. In addition, the factor ${2\o
\a^2}$ also indicates the corresponding interchange of the algebra and its
dual.

The one-soliton solutions of this model have a very simple structure. For each
generator $T_j(\a )$, ($j=1,2$), we have Sine-Gordon solitons excited
independently. However, if one-solitons of different Sine-Gordon models are
excited, they interact in a non-trivial way. The investigation of this last
point, which is in progress, is of great interest because such model describes
an integrable interaction of Sine-Gordon solitons.

\sect{Conclusions and discussion}
\label{sec:discuss}

\ \indent
We have constructed the Generalized Conformal Affine Toda models
(G-CAT), and studied the most salient aspects of their structure. Along
the text, we have remarked that most of these properties can be
viewed as the generalization of similar features of the Abelian
Conformal and of the Non-Abelian Affine Toda models. In our
opinion, one of the most important points is that the Non-Abelian Affine
Toda models can be understood as the result of the spontaneous breakdown
of the conformal symmetry of the G-CAT models. Consequently, the masses
of the solitons of the Affine Toda models are generated by a Higgs-like
mechanism. Even more, this approach allows one to put in one-to-one
correspondence the massive fundamental particles of the model with its
one-soliton solutions; a relation which we expect to be relevant when
quantizing these theories.

To study these G-CAT models, we have used different well known methods
in the field of the integrable non-linear systems, and we have
clarified the connection between all of them. Namely, in constructing the
general solutions of the model, we have used and, when needed,
we have suitably generalized the Leznov-Saveliev method for zero-curvature
equations~\ct{misha,misha2,misha3}, the dressing-transformation
approach~\ct{dress1,dress2,dress3,dress4}, and we have also defined the
appropriated
$\tau$-functions~\ct{hiro,wilson,tault}. The result is a systematic
procedure for constructing the one-soliton solutions and an
explicit formula for their masses.

Among the directions in which our work can be developed
further, we point out the following. First, as we have seen in
section~\ref{sec:symmetries}, some of the left and right translations on the
two-loop WZNW model remain as symmetries of the G-CAT  model. However, as it
happens in the abelian case, those which do not survive are twisted by the
reduction procedure and should give rise to non-linear symmetries, {\it
i.e.\/}, to $W$ algebras. This has already been investigated in the abelian CAT
models~\ct{higher,2boson,charges}. They have only two chiral remaining
Kac-Moody currents in each chiral sector that, in fact, can be used to
construct infinite $W$ algebras; the other symmetries do not lead to
chiral currents but to chiral charges~\ct{charges}. Moreover, it would
be interesting to investigate if some of the generators of the two-loop
Virasoro algebra constructed in section~\ref{sec:twovir} remain as
symmetries of the G-CAT model.

Second, a systematic study of the different G-CAT models according to
their specific properties has to be made. In particular, from the
point of view of their eventual quantization, it would be
particularly important to distinguish those models whose kinetic term is
positive-definite, and whose action is real. Results in this
direction have recently been presented in~\ct{tlp}, and, concerning the
first point, we have made some comments in sections~\ref{sec:reduction}
and~\ref{sec:symmetries}. In section~\ref{sec:symmetries}, we have
also pointed out that a family of models admitting static solutions
correspond to the choice of $E_{\pm l}$ as elements of some Heisenberg
subalgebra of $\cgh$. Then, the available classification of the
Heisenberg subalgebras of an affine Kac-Moody algebra~\ct{kacpet} can be
used to construct a large number of particular models whose specific
properties should be analysed.

Third, concerning the soliton solutions, their topological character
should be investigated to establish in which cases a topological
charge can be defined and also to understand its spectrum.

Finally, let us mention that the most interesting development to be
made is the construction of the quantum theory of these models, which
is expected to be integrable for a wide class of them. Then, it
should be possible to construct an exact factorizable S-matrix and to
compare the states of the resulting quantum theory with the solitons of
the classical one.

\vskip1cm

\begin{center}
{\large\bf Acknowledgements}
\end{center}

\vskip.2cm

\noindent
We are grateful to M.V.~Saveliev and A.V.~Ramallo for helpful
discussions. J.L.M. would like to thank Tim Hollowood for useful
conversations. L.A.F. was partially supported by Ministerio de
Educaci\'on y Ciencia (Spain) and FAPESP (Brazil). J.L.M. and J.S.G.
have been supported partly by CICYT (AEN93-0729) and DGICYT (PB90-0772)
(Spain).

\vskip.7cm

\appendix

\sect{Appendix: The Two-Loop Virasoro algebra}
\label{sec:twovir}

\ \indent
The Sugawara construction for the two-loop Kac-Moody algebra \rf{kma}-\rf{kmd}
can be used to obtain an algebra much bigger than the Virasoro algebra. In
order to do that we impose periodic boundary conditions on the currents.
Namely, we take the space variable $x$ being on a unit circle, and
require the currents to satisfy, $J_a^m(x+2\pi ) = J_a^m(x)$, $J^{{\cal
C}}(x+2\pi ) = J^{{\cal C}}(x)$ and $J^{{\cal D}}(x+2\pi ) = J^{{\cal D}}(x)$.
Since  $J^{{\cal C}}(x)$ commutes with $J_a^m(x)$ we can define a new set of
currents by \ct{schw}
\be
\tilde{J}^m_a(x) = J^m_a(x) exp\( {m\o {k}} \sum_{n \neq 0} {C_n \o
n}e^{- i n x}\) \lab{newcur}
\ee
where $C_n$ are the modes of $J^{{\cal C}}(x)$, {\it i.e.\/},
\be
J^{{\cal C}}(x) =  \sum_{n=-\infty}^{\infty} C_n e^{-inx}
\lab{expC}
\ee
The zero mode $C_0$ is not included in the transformation \rf{newcur} in
order to respect the periodicity properties of $\tilde{J}^m_a(x)$.
Notice that the transformation \rf{newcur} is defined on a given
representation and not on the abstract algebra. Using \rf{kma}-\rf{kmd} (with
the redefinitions $f_{ab}^c \ra if_{ab}^c$ and $k\ra ik$)  one obtains
\ct{schw}
\br
\lb \tilde{J}^m_a (x) \, , \, \tilde{J}^n_b (y) \rb &=&  i\, f_{ab}^c
\tilde{J}_c^{m+n} (x) \delta (x - y)  +  g_{ab}\delta_{m,-n} \(
i\, k\pa_x  \d (x - y)   +  m C_0  \d (x - y)  \)
\phantom{......}
\lab{kmta} \\
\lb D_0 \, , \, \tilde{J}^m_a (x) \rb &=& m \tilde{J}^m_a (x)
\lab{kmtb} \\
\lb D_n \, , \, \tilde{J}^m_a (x) \rb &=& 0 \quad\qquad\;\; n\neq 0
\lab{kmtc}
\er
where $D_n$ are the modes of $J^{{\cal D}}(x)$, {\it i.e.\/},    $J^{{\cal
D}}(x) = \sum_{n=-\infty}^{\infty} D_n e^{-inx}$. $D_0$ is the operator
measuring the ``momenta'' $m$ in an usual KM algebra \ct{go}. Formulae
\rf{kma}-\rf{kmc} show that the $J^{{\cal C}}$-$J^{{\cal D}}$ system is
decoupled from the $\tilde{J}'s$.  $C_0$ commutes with all the operators,
{\it i.e.\/}, acts like a second central extension in the algebra of
$\tilde{J}'s$.

Following the Sugawara construction we  define
\be
\cl_m(x) \equiv {1\o {2}} \sum_{a,b=1}^{{\rm dim}\, \lie}
\sum_{n=- \infty}^{\infty} g^{ab} {\tilde J}^{m+n}_{a} (x) {\tilde J}^{-n}_{b}
(x)
\lab{2lsuga}
\ee
Using \rf{kmta} one can check that
\br
\lb \cl_m(x)\, , \, \cl_n(y) \rb &=&
i\, k \( 2 \cl_{m+n}(y) \d^{\pr}(x-y) - (\pa_y\cl_{m+n}(y)) \d (x-y)\)
\nonu\\
&+& C_0 (m-n) \cl_{m+n}(y) \d (x-y)
\lab{2lvir}
\er

In addition, under $\cl_m(x)$ the currents transform as
\br
\lb \cl_m(x)\, , \, {\tilde J}^{n}_{a}(y) \rb &=&
i\, k \( {\tilde J}^{m+n}_{a}(y) \d^{\pr} (x-y) - \pa_y\( {\tilde
J}^{m+n}_{a}(y)\) \d (x-y) \) \nonu\\
&-& n \, C_0 \, {\tilde J}^{m+n}_{a}(y) \d (x-y)
\lab{2lvircur}
\er

Since the currents $\tilde{J}^m_a(x)$ satisfy periodic boundary conditions, so
does $\cl_m(x)$. We then expand it in modes as
\be
\cl_{m_1}(x) \equiv \sum_{m_2=-\infty}^{\infty} \cl_{(m_1,m_2)}
e^{-im_2 x}
\ee
Using  \rf{2lvir} one gets
\be
\lb \cl_{(m_1,m_2)} \, , \, \cl_{(n_1,n_2)}\rb = \( C_0 \, (m_1-n_1) +
k \, (m_2-n_2)\)
\cl_{(m_1+n_1,m_2+n_2)}
\lab{2lvird}
\ee
where we have expanded the delta function as $\d (x-y) =
\sum_{n=-\infty}^{\infty}e^{-in (x-y)}$.

If the central terms $C_0$ and $k$ are rational numbers, then one can always
eliminate one of them from the above relation  by a redefinition of the
integer labels of $\cl_{(m_1,m_2)}$. We introduce the $sl(2, \IZ )$
transformation
\br
\( \begin{array}{c}
m_1^{\pr}\\
m_2^{\pr}
\end{array}\) = M \,
\( \begin{array}{c}
m_1\\
m_2
\end{array}\)
\lab{sl2z}
\er
where $M= \( \begin{array}{cc}
a&b\\
c&d
\end{array}\)$;
with $a$, $b$, $c$, $d$ $\in \IZ$ and ${\rm det} M =1$. The last condition is
needed in order to make the transformation one to one. We then define the new
generators
\be
l_{(m_1,m_2)} \equiv \cl_{(m_1^{\pr},m_2^{\pr})}
\ee
{}From \rf{2lvird} and \rf{sl2z} we get
\be
\lb l_{(m_1,m_2)} \, , \, l_{(n_1,n_2)}\rb = \( C_0^{\pr} \, (m_1-n_1) +
k^{\pr} \, (m_2-n_2)\)
l_{(m_1+n_1,m_2+n_2)}
\ee
where
\be
\( C_0^{\pr} \, k^{\pr} \) = \( C_0 \, k \) M
\ee
Now, if the central terms are rational, let us say $C_0 = p/q$ and $k = {\bar
p}/{\bar q}$, then we can choose $a=q{\bar p}$ and $c=-p{\bar q}$, and $b$ and
$c$ such that $d q{\bar p} + b p{\bar q} =1$, which always has a solution.
Therefore we get $C_0^{\pr} =0$ and $k^{\pr}=1/q{\bar q}$.

Expanding the currents ${\tilde J}^{m}_{a}(x)$ in $x$, and making similar
redefinitions in the pair of indices of its modes one can eliminate one of the
central terms in \rf{kmta} \ct{schw}.

When the finite Lie algebra associated to the structure constants $f_{ab}^c$ is
compact, then it can be shown \ct{schw} that for unitary, highest weight
representations for the currents $\tilde{J}^m_a (x)$ the  central terms $k$
and $C_0$ have to be integers \ct{schw}. In such a case the
elimination procedure above works. The reduction procedure discussed in
section~\ref{sec:reduction} uses the Gauss decomposition. Therefore  the
two-loop Kac-Moody algebra \rf{kma}-\rf{kmd} has to be based on a finite
simple {\em non-compact} Lie algebra. In such cases $k$ and
$C_0$ are not necessarily integers and can in principle assume any value
(even complex).

\sect{Appendix: Affine Kac-Moody algebras}
\label{ap:km}

\ \indent
In the Chevalley basis, the commutation relations for an untwisted affine
Kac-Moody algebra $\cgh$ are
\br
\lb H_a^m \, , \,  H_b^n \rb &=& m\, C \, \eta_{ab}  \,\d_{m+n,0}
\lab{km1}\\
\lb H_a^m \, , \,  E_{\a}^n \rb &=& \sum_{b=1}^r m_{b}^{\alpha} K_{ba}
E_{\a}^{m+n}
\lab{km2}\\
\lb E_{\a}^m \, , \,  E_{-\a}^n \rb &=& \sum_{a=1}^r l_{a}^{\alpha} \,
H^{m+n}_a    +  {2\o \a^2}\, m \, C \, \d_{m+n,0}
\lab{km3}\\
\lb E_{\a}^m \, , \,  E_{\b}^n \rb &=&  (q+1)\, \epsilon (\a , \b )\, E_{\a
+\b}^{m+n} \, \, ;
\qquad \qquad \mbox{\rm if $\a + \b$ is a root}
\lab{km4}\\
\lb D \, , \,  E_{\a}^m \rb &=& m \, E_{\a}^m
\lab{km5}\\
\lb D \, , \,  H_a^m \rb &=& m \, H_a^m
\lab{km6}
\er
where $K_{ab}= 2 \a_a \cdot \a_b / \a_b^2$ is the Cartan matrix of the finite
simple Lie algebra $\cg$ associated to $\cgh$ and generated by $\{H_{a}^0,
E_{\alpha}^0\}$. $\eta_{ab}= {2\o
\a_a^2}\, K_{ab} = \eta_{ba}$, $q$ is the highest positive integer such that
$\b - q \a$ is a root,
$\epsilon (\a , \b )$ are signs determined by the Jacobi identities,
$l_{a}^{\alpha}$ and
$m_{a}^{\alpha}$ are the integers in the expansions $\a /\a^2 = \sum_{a=1}^r
l_{a}^{\alpha}\> \a_a / \a_a^2$ and  $\a  = \sum_{a=1}^r m_{a}^{\alpha}\> \a_a
$, and $\a_1,\ldots,\a_r$ are the simple roots of $\cg$ ($r \equiv
\mbox{\rm rank $\cg$}$).
$\cgh$ has a symmetric non-degenerate bilinear form of $\cgh$ which can be
normalized as
\br
\Tr \( H_a^m \, H_b^n \) &=& \eta_{ab} \d_{m+n,0}
\lab{trace1}\\
\Tr \( E_{\a}^m \,  E_{\b}^n \) &=& {2\o \a^2} \d_{\a + \b ,0} \, \d_{m+n,0}
\lab{trace2}\\
\Tr \( C \, D \) &=& 1
\lab{trace3}
\er

The integral gradations of $\cgh$
\be
\cgh = \bigoplus_{n \,\in \IZ} \cgh_n
\ee
have been classified in \ct{kac1,kac2}; the result is that every integer
gradation with finite dimensional $\cgh_0$ is conjugate to a gradation defined
by a grading operator $\qs$ satisfying
\be
\lb Q_{\bf s} \, , \, \cgh_n \rb = n \, \cgh_n \, \, ; \qquad n \in \IZ,
\lab{niceq}
\ee
and defined by
\be
Q_{\bf s} = H_{\bf s} + N_{\bf s}\, D + \sigma C,\quad H_{\bf s}=
\sum_{a=1}^{r} s_a \l^v_a \cdot H^0,
\lab{gradop}
\ee
where $(s_0, s_1, \cdots ,s_r)$ is a vector of non-negative
relatively prime  integers, and $\l^v_a \equiv 2 \lambda_a/\alpha_a^2$ with
$\lambda_a$ and $\alpha_a$ being the fundamental weights and simple roots of
$\cg$ respectively. Moreover,
\be
N_{\bf s} = \sum_{i=0}^{r} s_i \, m_i^{\psi}, \qquad
\psi = \sum_{a=1}^{r}m_a^{\psi}\,  \alpha_a , \qquad
m_0^{\psi} \equiv 1
\lab{grad}
\ee
with $\psi$ being the maximal root of $\cg$; obviously, the value of $\sigma$
is
arbitrary, but we shall make the standard choice $\sigma= -Tr(H_{\bf
s}^2)/(2N_{\bf s})$, which ensures that $Tr(Q_{\bf s}^2)=0$. Two gradations
are   equivalent if the corresponding vectors $(s_0,s_1,\cdots ,s_r)$ and
$(s'_0, s'_1, \cdots s'_r)$ are related by a symmetry of the extended Dynkin
diagram of $\cgh$. Therefore we have
\br
\lb Q_{\bf s} \, , \, H_a^n \rb &=& n N_{\bf s} \, H_a^n
\lab{grad1}\\
\lb Q_{\bf s} \, , \, E_{\a}^n \rb &=& \( \sum_{a=1}^r m_a s_a + n N_{\bf s}
\)\,  E_{\a}^n
\lab{grad2}
\er
The positive and negative simple root step operators of $\cgh$ are $e_a\equiv
E_{\a_a}^0$, $e_0 \equiv E_{-\psi}^1$ and $f_a\equiv E_{-\a_a}^0$, $f_0 \equiv
E_{\psi}^{-1}$, and its Cartan subalgebra generators
$h_a \equiv H_a^0$, $h_0 \equiv - \sum_{a=1}^r l_a^{\psi} H_a^0 + {2\o
\psi^2}\, C$, and $D$, with $l_a^{\psi}$ given in \rf{lipsi}\footnotemark
{\footnotetext{The positive integer numbers $a_i= m_{i}^{\psi}$ and
$a_{i}^{\vee}= l_{i}^{\psi}\, \psi^2/2$, where $i=0,1,\ldots,r$,  are the
Kac-labels and the dual Kac-labels of $\cgh$, respectively~\ct{kac1}, and
$a_i = a_{i}^{\vee}\, \psi^2/\alpha_{i}^{2}$}}; then, they satisfy
\be
\lb Q_{\bf s} \, , \, h_i \rb = \lb Q_{\bf s} \, , \, D \rb =0 \, \, ;\quad
\lb Q_{\bf s} \, , \, e_i \rb =  s_i \, e_i \, \, ; \quad
\lb Q_{\bf s} \, , \, f_i \rb = - s_i \, f_i \, \, ;\qquad i=0,1,\ldots, r.
\lab{gradesimple}
\ee
A significant consequence of the existence of a grading operator is that
$\Tr(\cgh_j\, \cgh_k)=0$ if $j+k\not=0$, and that the subspaces $\cgh_j$ and
$\cgh_{-j}$ are non-degenerately paired for all $j\in \IZ$.

An important class of representations of the Kac-Moody algebras are the so
called {\em integrable highest weight representations} \ct{kac1}. They are
defined in terms of a highest weight state $\mid \l_{\bf s} \rangle$ labelled
by a gradation ${\bf s}$ of $\cgh$ \rf{niceq}. Such  state is annihilated by
the
positive grade generators
\be
\cgh_{+} \mid \l_{\bf s} \rangle = 0
\lab{int0}
\ee
and it is an eigenstate of all generators of the subalgebra $\cgh_0$
\br
h_i \mid \l_{\bf s} \rangle &=& s_i \mid \l_{\bf s} \rangle
\lab{int1}\\
f_i \mid \l_{\bf s} \rangle &=& 0\, \, ; \qquad \mbox{\rm for any $i$ such that
$s_i =0$}
\lab{int2}\\
Q_{\bf s} \mid \l_{\bf s} \rangle &=& \eta_{\bf s}\mid\l_{\bf s}\rangle
\lab{int3}\\
C \mid \l_{\bf s} \rangle &=& {\psi^2\o 2}\, \( \sum_{i=0}^r l_i^{\psi} s_i\)
\mid
\l_{\bf s}
\rangle
\lab{int4}
\er
where $l_i^{\psi}$ given by
\be
{\psi\o \psi^2} = \sum_{a=1}^r l_a^{\psi} {\a_a \o \a_a^2} \, \, ; \qquad
l_0^{\psi} = 1.
\lab{lipsi}
\ee
It is always possible to modify the definition of the grading operator $Q_{\bf
s}$ by adding a component in $C$ in such a way that the eigenvalue $\eta_{\bf
s}$ vanishes.

The integers $m_i^{\psi}$ and $l_i^{\psi}$, given in \rf{grad} and \rf{lipsi},
constitute respectively the left and right null vectors of the extended Cartan
matrix of $\cgh$
\be
\sum_{i=0}^r m_i^{\psi}K_{ij} =0 \qquad \qquad
\sum_{j=0}^r K_{ij} l_j^{\psi}=0;
\ee
notice that $[h_i,e_j]=K_{ji} e_j$ for all $i,j=0,1,\ldots,r$.

The highest weight states $\mid \l_{\bf s} \rangle$ can be realized as
\be
\mid \l_{\bf s} \rangle \equiv \bigotimes_{i=0}^r \mid {\hat
\l}_{i}\, \rangle^{\otimes s_i}
\lab{lambdas}
\ee
where $\mid {\hat \l}_{i}\, \rangle$ are the highest weight states of the
fundamental representations of $\cgh$, and ${\hat \l}_{i}$ are the
corresponding fundamental weights of $\cgh$. Namely \ct{go}
\br
{\hat \l}_{0} &=& (0,\psi^2/2,0)
\lab{lambda0} \\
{\hat \l}_{a} &=& ( \l_a , l_a^{\psi} \psi^2/2,0)
\lab{lambdaa}
\er
where $\l_a$, $a=1,2, \ldots r$ are the fundamental weights of the finite Lie
algebra $\cg$ associated to $\cgh$, $l_a^{\psi}$ is defined in
\rf{lipsi}, and the entries are the eigenvalues of $H_a^0$, $C$ and $D$
respectively, {\it i.e.\/},
\br
H_a^0 \, \mid {\hat \l}_{0}\, \rangle &=& 0 \, ; \qquad
C \, \mid {\hat \l}_{0}\, \rangle = {\psi^2 \o 2} \, \mid {\hat \l}_{0}\,
\rangle
\lab{fundrep0}\\
H_b^0 \, \mid {\hat \l}_{a}\, \rangle &=& \d_{a,b}\mid {\hat \l}_{a}\,
\rangle   \, ; \qquad
C \, \mid {\hat \l}_{a}\, \rangle = {\psi^2 \o 2}l_a^{\psi} \,
\mid {\hat \l}_{a}\,\rangle
\lab{fundrepa}
\er
and
\be
D \, \mid {\hat \l}_{i}\, \rangle = 0
\ee

Notice that in each of the $r+1$ fundamental representations of $\cgh$, the
(unique) highest weight state satisfies
\br
h_j \, \mid {\hat \l}_{i}\, \rangle &=& \d_{ij} \mid {\hat \l}_{i}\, \rangle
\\
e_j \, \mid {\hat \l}_{i}\, \rangle &=& 0
\, ; \qquad \mbox{\rm for any $j$}\\
f_j \, \mid {\hat \l}_{i}\, \rangle &=& 0
\, ; \qquad \mbox{\rm for $j\neq i$}\\
f_i^2 \, \mid {\hat \l}_{i}\, \rangle &=& 0.
\er
Therefore the generators $e_i$ and $f_i$ are nilpotent when acting on
\rf{lambdas}, and these representations are actually integrable.

\sect{Appendix: A particular basis of $\cgh_0$}
\label{ap:gcero}

\ \indent
Let us consider the integer gradation associated to a vector
$(s_0,s_1,\ldots,s_r)$ defined by the grading operator
\be
Q_{\bf s}\, =\, H_{\bf s}\, +\, N_{\bf s}\, D\, -\, {1\over 2 N_{\bf
s}}\,Tr(H_{\bf s}^2)\, C,
\lab{choicegrad}
\ee
where
\be
H_{\bf s} = \sum_{a=1}^{r} s_a \, \lambda_{a}^{v}\cdot H^0,
\ee
according to eq.\rf{gradop}. $H_{\bf s}$ is an element of the Cartan
subalgebra of the finite Lie algebra $\cg$ (that one with zero grade w.r.t.
$D$), and it defines a finite integer gradation of $\cg$
\be
\cg\, =\, \bigoplus_{j=-L}^{+L} \, \cg_{j},\qquad
[H_{\bf s}\,,\,v]\,= \, j\,v \quad {\rm for\;\; all}\quad v\in \cg_{j}.
\lab{liegrad}
\ee
Notice that the step operator corresponding to the maximal root has to be an
element of maximal grade, {\it i.e.\/}, $E_\psi\in \cg_L$; but the grade  of
$E_\psi$ can be obtained using eq.\rf{grad2}, which, taking into account
\rf{grad} too, implies that
\be
L\, =\, \sum_{a=1}^{r}\, s_a \,m_{a}^{\psi}\, =\, N_{\bf s} \, -\, s_0.
\ee
Therefore, in addition to $C$ and $D$, the zero graded subspace of $\cgh$
will contain $\cg_{0}^0$ when $s_0\not=0$, or $\cg_{0}^0\oplus \cg_{N_{\bf
s}}^{-1} \oplus \cg_{-N_{\bf s}}^{+1}$ when $s_0=0$; notice that we have
introduced the notation $\cg_{k}^m =\{u^m\in \cgh\bigm|\,  u\in \cg_k\}$.

In the case where $s_0 = 0$, the commutator of two elements of $\cgh_0$ may
produce terms in the direction of the central term $C$.
Therefore, in order to establish the relation between the Conformal Affine
(G-CAT) and the Affine (G-AT) Toda models, it is convenient to choose a basis
of the subspace $\cgh_0$ in which the affine components $\cg_{0}^0$ and
$\cg_{\pm N_{\bf s}}^{\mp1}$ are orthogonal to the other components of
$\cgh_0$; this only requires to change the basis of the subalgebra of $\cgh_0$
generated by
$C$, $D$, and the Cartan subalgebra of $\cg_0$. Taking into account the
eqs.\rf{trace1}-\rf{trace3}, a convenient choice is
$\{C, Q_{\bf s},\widetilde{H}_{1}^0,\ldots,\widetilde{H}_{r}^0\}$, where
\br
\widetilde{H}_{a}^0\, &=&\, H_{a}^0\, -\, {1\over N_{\bf s}}\, \Tr\left(H_{\bf
s}\, H_{a}^0\right)\, C\nonu\\
&=&\, H_{a}^0\, -\, {2\over \alpha_{a}^2}{s_a\over N_{\bf s}}\, C,
\lab{cartanplus}
\er
which satisfies
\br
&&\Tr\left(C^2\right)= \Tr\left(C\,\widetilde{H}_{a}^0\right) = \Tr\left(
Q_{\bf s}^2 \right)=
\Tr\left( Q_{\bf s}\, \widetilde{H}_{a}^0\right)=0,
\lab{bilbasis1}\\
&&\Tr\left( Q_{\bf s}\, C\right) = N_{\bf s},\qquad \Tr\left(
\widetilde{H}_{a}^0\, \widetilde{H}_{b}^0\right)= \Tr\left(H_{a}^0\,
H_{b}^0\right) = \eta_{ab},
\lab{bilbasis2}
\er
for all $a,b=1\ldots,r$.

Now, let $\cgh_{0}^\ast$ be the subspace of $\cgh_0$ generated by
$\{\widetilde{H}_{1}^0, \ldots, \widetilde{H}_{r}^0\}$ and by either
$\{E_{\alpha}^0 \bigm|\, E_{\alpha}\in \cg_0\}$, if $s_0\not=0$, or
$\{E_{\alpha}^0, E_{\pm\beta}^{\mp1}\bigm|\, E_\alpha\in \cg_0,\;
E_{\pm\beta}\in \cg_{\pm N_{\bf s}}\}$, if $s_0=0$. Then, $\cgh_{0}^\ast$ is a
subalgebra of $\cgh_0$, and it is isomorphic either to $\cg_{0}^0$, when
$s_0\not=0$, or to $\cg_{0}^0\oplus \cg_{-N_{\bf s}}^{1} \oplus \cg_{+N_{\bf
s}}^{-1}$, if $s_0=0$. Moreover, by construction, $\cgh_{0}^\ast$ commutes with
$Q_{\bf s}$ and $C$; therefore, any element $u\in \cgh_0$ can be
parameterized as
\be
u\, = \, u^{\ast}\, +\, \nu\,C \,+\, \eta \,Q_{\bf s},\quad
u^{\ast}\in \cgh^{\ast}_0,
\lab{paramalg}
\ee
where
\be
\nu ={1\over N_{\bf s}} \Tr\left( u\> \qs\right),\qquad
\eta = {1\over N_{\bf s}} \Tr\left( u\> C\right).
\lab{projection}
\ee
Correspondingly, any element of the group formed by exponentiating the
elements of $\cgh_0$ can be expressed as
\be
B= B_{0} \, \exp\left(\nu\, C\, + \,\eta\, Q_{\bf
s}\right),
\lab{paramgroup}
\ee
where $B_{0}$ is an element of the subgroup obtained by
exponentiating the subalgebra $\cgh_{0}^{\ast}$. It is worth noticing that
trace form $\Tr$, restricted to $\cgh_{0}^{\ast}$, is also non-degenerate.
Finally, let us point out that, when $s_0\not=0$,
$B_{0}$ is the field of the non-abelian affine Toda model associated to
the affine Lie algebra $\cgh$ and to the gradation induced by $Q_{\bf s}$.
However, when $s_0=0$, we have models of non-abelian affine type which, as
far as we know, were not considered in the literature before.

In order to substitute eq.\rf{paramgroup} in eq.~\rf{brokenwznw}, the following
expressions will be useful:
\br
&& \Tr\left(\partial_\mu B\> \partial^\mu B^{-1}\right) =
\Tr\left(\partial_\mu B_0\> \partial^\mu B_{0}^{-1}\right) \,-\, 2\>N_{\bf s}\>
\partial_\mu \nu\> \partial^\mu \eta,\lab{paramWZNW1}\\
&& \epsilon^{ijk} \Tr\left(B^{-1}\partial_i B\> B^{-1}\partial_j B\>
B^{-1}\partial_k B\right) = \epsilon^{ijk} \Tr\left(B_{0}^{-1}\partial_i
B_{0}\>
B_{0}^{-1}\partial_j B_{0}\> B_{0}^{-1}\partial_k B_{0}\right)
,\lab{paramWZNW2}\\
&& \Tr\left(\Lambda_l\> B^{-1}\> \Lambda_{-l}\> B\right) =
e^{l\>\eta}\,
\Tr\left(\Lambda_l\> B_{0}^{-1}\> \Lambda_{-l}\> B_{0}\right).
\lab{paramWZNW3}
\er

\end{document}